\documentclass[twocolumn,amsmath,amssymb,prd]{revtex4-1} 
\usepackage{graphicx}
\usepackage{pstricks}
\usepackage{textcomp}
\usepackage{multirow}

\def\al{\alpha}
\def\be{\beta}
\def\ga{\gamma}
\def\de{\delta}
\def\ep{\epsilon}

\def\et{\eta}

\def\ka{\kappa}
\def\la{\lambda}

\def\rh{\rho}

\def\si{\sigma}

\def\ph{\phi}

\def\ch{\chi}
\def\ps{\psi}
\def\om{\omega}

\def\De{\Delta}

\def\La{\Lambda}
\def\Si{\Sigma}

\def\at{{\widetilde a}}
\def\bt{{\widetilde b}}
\def\ct{{\widetilde c}\hskip1pt}
\def\dt{{\widetilde d}\hskip2pt}
\def\gt{{\widetilde g}}
\def\Ht{{\widetilde H}}

\def\dash{\multicolumn{1}{c}{--}}

\def\pr{^\prime}

\def\hc{\textrm{h.c.}}

\def\cL{{\cal L}}

\def\half{{\textstyle{\frac 1 2}}}
\def\quar{{\textstyle{\frac 1 4}}}
\def\eigh{{\textstyle{\frac 1 8}}}

\def\lsim{\mathrel{\rlap{\lower4pt\hbox{\hskip1pt$\sim$}}
    \raise1pt\hbox{$<$}}}
\def\gsim{\mathrel{\rlap{\lower4pt\hbox{\hskip1pt$\sim$}}
    \raise1pt\hbox{$>$}}}

\def\prt{\partial}

\newcommand{\beq}{\begin{equation}}
\newcommand{\eeq}{\end{equation}}
\newcommand{\bea}{\begin{eqnarray}}
\newcommand{\eea}{\end{eqnarray}}
\newcommand{\rf}[1]{(\ref{#1})}
\newcommand{\nn}{\nonumber}

\def\etal{{\it et al.}}

\def\psb{\overline{\ps}{}}

\def\ov{\overline}

\def\ring#1{{\mathaccent'27 #1}}

\def\eff{\textrm{asy}}
\def\NR{\textrm{NR}}
\def\wb{{\ol{w}}}

\def\L{\textrm{L}}

\def\ol#1{\overline{#1}{}}
\def\vev#1{\langle {#1}\rangle}
\def\abs#1{\vert{#1}\vert}

\def\aaa{ss}
\def\bbb{\Si\Si}

\def\atom{{\hskip 2pt\rm atom}}
\def\rel{{\hskip 2pt\rm rel}}
\def\latom{{\rm atom}}

\def\vev#1{\langle {#1}\rangle}

\def\bra#1{\langle{#1}|}
\def\ket#1{|{#1}\rangle}

\def\kk{{i}}
\def\kkk{{i^\prime}}

\def\xx{{x^\prime}}
\def\zz{{z^\prime}}

\begin{document}

\title{
Searches for beyond-Riemann gravity}

\author{
V.\ Alan Kosteleck\'y and Zonghao Li}

\affiliation{
Physics Department, Indiana University, 
Bloomington, IN 47405, USA}

\date{June 2021; accepted for publication in Physical Review D} 

\begin{abstract}

Many effective field theories describing gravity 
cannot arise from an underlying theory based on Riemann geometry
or its extensions to include torsion and nonmetricity
but may instead emerge from another geometry
or may have a nongeometric basis.
The Lagrange density for a broad class of such theories is investigated.
The action for fermions coupled to gravity 
is linearized about a Minkowski background
and is found to include terms 
describing small deviations from Lorentz invariance
and gravitational gauge invariance.
The corresponding nonrelativistic hamiltonian is derived
at second order in the fermion momentum.
The implications for laboratory experiments and astrophysical observations
with fermions are studied,
with primary focus on anomalous spin-gravity couplings.
First constraints on some coefficients 
are extracted from existing data 
obtained via measurements at different potentials, 
comparisons of gravitational accelerations,
interferometric methods,
and investigations of gravitational bound states.
Some prospects for future experimental studies are discussed.

\end{abstract}

\maketitle

\section{Introduction}
\label{Introduction}

The construction of a compelling and realistic underlying theory 
that unifies gravity with quantum physics remains an open challenge.
The description of spacetime in the underlying theory might involve 
the usual Riemann geometry of General Relativity (GR),
or it might be a non-Riemann geometry or have no geometrical basis.
An interesting and potentially vital issue 
is then the extent to which current or feasible experiments
can help to distinguish between these possibilities.

Coupling GR to the Standard Model (SM) of particle physics
produces a theory that is quantum incomplete 
but that yields an excellent match to experiments in suitable regimes.
Any deviations from known physics emerging from the underlying unified theory
are therefore expected to be small,
perhaps suppressed by a large scale such as the Planck mass.
A model-independent approach to studying small deviations from a known theory
is provided by effective field theory (EFT)
\cite{sw09}.
Investigations of the geometric properties 
of the underlying unified theory can therefore be based 
on the general EFT constructed from the action of GR coupled to the SM.

To allow for deviations from Riemann geometry,
the general EFT must contain 
both terms preserving and violating the spacetime symmetries of GR,
which include the invariances 
under local Lorentz transformations and diffeomorphisms.
The general EFT based on GR coupled to the SM
is presented in Ref.\ \cite{ak04}.
Each additional term in the action 
involves a coupling coefficient 
combined with an operator constructed from dynamical fields.
A given coupling coefficient $k$ can be viewed as a background
that can carry spacetime or local indices
and can depend on spacetime position.
Except for the special case of a constant scalar coupling,
any coefficient controls violations of one ore more of the invariances of GR.
The coefficients can also be flavor dependent,
so violations of the weak equivalence principle (WEP) are incorporated. 
A given term in the EFT action can be classified
according to the mass dimension $d$ of the dynamical operator it contains,
with minimal terms defined to have $d\leq 4$ and nonminimal ones $d\geq 5$.
A systematic construction of all terms has recently been presented
in Ref.\ \cite{kl21}.

The background coefficients can be dynamical or prescribed quantities.
In the former case they are called spontaneous,
and in the latter explicit.
At the EFT level,
a spontaneous background $k = \vev{k} + \de k$
consists of a vacuum value $\vev{k}$
solving the equations of motion,
together with dynamical fluctuations $\de k$ about $\vev{k}$
that include Nambu-Goldstone and massive modes.
In contrast,
an explicit background $k = \ol k$ is a predetermined quantity.
The presence of the fluctuations $\de k$
generates corresponding physical effects,
which can distinguish a spontaneous background $\vev{k}$ 
from an explicit background $\ol k$.
In the present work,
we focus primarily on EFT based on GR coupled to the SM
and containing one or more explicit backgrounds $\ol k$.

The structure of an EFT with an explicit background $\ol k$ 
can be constrained by the requirement of compatibility
between the variational procedure 
and the Bianchi identities of Riemann geometry
\cite{ak04,rb15,kl21}.
It turns out that most EFT terms with explicit backgrounds 
are perturbatively incompatible with Riemann geometry
or its extensions with torsion and nonmetricity,
and hence typical models containing terms of this type
must be based on a different geometry or have a nongeometric origin. 
These no-go constraints provide a powerful tool
to specify terms that cannot arise in Riemann geometry
and thereby to identify physical effects serving as experimental signals 
for an underlying unified theory based on nonstandard geometry
or on a nongeometrical structure.

In the present work,
we explore this line of reasoning 
by investigating physical effects in a large class of EFT 
based on GR coupled to the SM
and incorporating terms that involve beyond-Riemann effects.
For this initial study,
we focus on gravitational couplings of fermions
and in particular spin-gravity couplings,
which have implications for many existing 
laboratory and astrophysical observations.
We consider in turn experiments involving
measurements at different potentials,
comparisons of gravitational accelerations,
interferometric methods,
and gravitational bound states.
We use existing experimental data
to obtain first constraints on the EFT coefficients
governing beyond-Riemann physics.
We also discuss prospects for some future experimental studies.

The organization of this paper is as follows.
The setup for the EFT containing beyond-Riemann effects
is presented in Sec.\ \ref{Theory}.
We provide tables detailing the terms in the action
and their linearizations.
The corresponding nonrelativistic hamiltonian is obtained,
and its coefficients are related
to those in the linearized Lagrange density.
The flavor dependence of the coefficients
corresponding to WEP violations is discussed,
including implications for antiparticles.
This material makes feasible the analysis of various experiments.
In Sec.\ \ref{Potential differences}
we adopt results from existing experiments performed at different potentials 
to extract first constraints on some coefficients in the EFT.
Another class of experiments,
analyzed in Sec.\ \ref{Free-fall experiments},
involves comparing the gravitational accelerations of different atoms.
We consider tests with Sr atoms of different spins
and with Rb atoms in different hyperfine states,
and we discuss the prospects for comparing 
the gravitational accelerations of matter and antimatter.
In Sec.\ \ref{Interferometer experiments},
we turn attention to interferometric experiments with neutrons
to obtain sensitivity to additional coefficients.
Studies of neutrons bound in the Earth's gravitational field
can also provide interesting measurements
and are treated in Sec.\ \ref{Bound-state Experiments}.
We summarize our results in Sec.\ \ref{Discussion}.

Throughout this work,
we follow the conventions of Ref.\ \cite{kl21},
using natural units with $c=\hbar=\ep_0=1$.
For the experimental analyses, 
we adopt standard reference frames widely used in the literature.
No laboratory on the Earth lies in an inertial frame,
so experimental results for coefficients 
are reported in the canonical Sun-centered frame
\cite{sunframe}.
This right-handed orthogonal system
has spatial coordinates $J=X,Y,Z$,
with the $Z$ axis aligned along the Earth's rotation axis
and with the $X$ axis pointing from the Earth to the Sun 
at the vernal equinox 2000,
which serves as the zero of the time $T$.
For some calculations,
it is convenient to adopt a canonical laboratory frame
having time coordinate $t$ and spatial coordinates $j=x,y,z$,
with the $z$ axis oriented toward the local zenith
\cite{sunframe}.
Neglecting the Earth's boost,
the transformation from the Sun-centered frame 
to the canonical laboratory frame is given by the rotation matrix
\beq
R^{jJ}= \left(
\begin{array}{ccc}
\cos\ch \cos\om_\oplus T_\oplus & \cos\ch \sin\om_\oplus T_\oplus & -\sin\ch
\\
-\sin\om_\oplus T_\oplus & \cos \om_\oplus T_\oplus & 0 
\\
\sin \ch \cos \om_\oplus T_\oplus & \sin \ch \sin\om_\oplus T_\oplus 
& \cos \ch
\end{array} \right),
\label{eq:rotation}
\eeq
where $\ch$ is the laboratory colatitude,
$\om_\oplus\simeq 2\pi/(23 \textrm{h } 56 \textrm{m})$ 
is the sidereal frequency of the Earth's rotation,
and $T_\oplus$ is a suitable local sidereal time.
In what follows,
some expressions involve coefficients with indices
either summed over $tt$, $xx$, $yy$, $zz$
and denoted for brevity by the pair $\aaa$,
or summed over $TT$, $XX$, $YY$, $ZZ$
and denoted by the pair $\bbb$.

\section{Theory}
\label{Theory}

One goal of this work is to construct an EFT
based on GR coupled to the SM
that contains a general class of terms excluded in Riemann geometry.
This provides a window on physics effects from beyond-Riemann theories
and permits the extraction of experimental constraints.
Here, 
we focus specifically on fermion-gravity couplings,
which are ubiquitous and comparatively straightforward to analyze
for most laboratory experiments and astrophysical observations,
while maintaining a broad and model-independent perspective.

Our primary interest lies in spin-gravity couplings,
in part because they are particularly challenging to fit into Riemann geometry
\cite{kl21}, 
but in this section we include effects from spin-independent terms as well.
The methodology presented here could therefore be applied
to measurements of spin-independent fermion-gravity couplings,
including precision experiments associated with WEP tests
\cite{ti20}. 
Both spin-dependent and spin-independent fermion-gravity couplings 
are known to arise in beyond-Riemann contexts.
In Finsler geometry
\cite{rf,bcs00},
for example,
the metric is supplemented with objects on the manifold 
that have been conjectured to play the role 
of the explicit backgrounds $\ol k$ 
in generic EFT based on GR coupled to the SM
\cite{ak04}.
Although a complete demonstration of this link 
awaits the resolution of open issues in Lorentz-Finsler geometry
\cite{ak11,bjs20,ps18,em17,tbgm18,ilp18,%
cs18,rt18,xl18,sma19,sv20,rl20,fhpv20},
the resulting trajectories arising from spin-gravity couplings
in a fixed gravitational background
are known to correspond to Riemann-Finsler geodesics
\cite{ak11,ms19,dc17,fl15,nr15}.
Possible gravitational couplings to boson fields, 
including photons,
are also of definite interest but lie beyond our present scope.

\subsection{Setup}
\label{Setup}

In theories with a geometric foundation, 
the equations of motion obtained from the variational principle 
are supplemented by geometric conditions
called the Bianchi identities,
which arise from the structure of the corresponding fiber bundle
\cite{mtw}.
For many geometric theories,
such as Maxwell electrodynamics 
or other gauge theories in Minkowski spacetime, 
the Bianchi identities are homogeneous equations
that are independent of the inhomogeneous equations
generated via the variational procedure. 
In contrast,
in some geometric theories such as GR,
the equations of motion are entangled with the geometry. 
The Bianchi identities then impose a self-consistency condition,
which for GR turns out to be the requirement 
of covariant conservation of the energy-momentum tensor. 
This is compatible with the matter equations of motion, 
so GR is self consistent.
However,
for some theories the geometric conditions can be incompatible 
with the results of the variational principle
and hence serve as no-go constraints
\cite{ak04}. 

This issue is of particular relevance for theories 
that purport to be based on Riemann geometry
but that violate spacetime symmetries. 
The geometric constraints become particularly stringent for theories
required to produce only perturbative corrections to GR at low energies 
while also maintaining the structure of Riemann geometry. 
Indeed,
most theories of this type with explicit breaking of spacetime
symmetries are incompatible with Riemann geometry
\cite{ak04}.
Many examples illustrating the no-go constraints are known
\cite{ak04,kl21,rb15}. 
The simplest may be the extension of GR containing a cosmological term 
that involves a prescribed nontrivial function $\ol\La(x)$ 
of spacetime position
\cite{rb15}, 
resulting in explicit diffeomorphism violation (EDV).
Variation of the action yields Einstein equations taking the usual form, 
but these are incompatible with the Bianchi identities
unless $\ol\La(x)$ is a spacetime constant,
which contradicts the initial assumption. 
A recent general discussion of the no-go constraints 
along with other examples is given in Sec. II F of Ref.\ \cite{kl21}.

The present work is based on the above results.
We take advantage of the potential incompatibility 
between the Bianchi identities and the variation of the action 
to investigate a large class of possible underlying theories 
that have a non-Riemann geometry or a nongeometric basis
while nonetheless reducing at low energies 
to a perturbatively corrected version of GR coupled to the SM. 
The perturbative nature implies that this class of theories
can naturally be studied in the EFT framework
\cite{sw09}. 
A typical application of the framework would involve
constructing a specific EFT 
based on integrating over high-energy degrees of freedom in a theory 
and ensuring self-consistency via loops.
Here, however,
we adopt a different approach, 
designed instead to study simultaneously 
a large class of theories with a given symmetry structure 
and to investigate their possible phenomenological EFT signatures. 
The approach involves constructing all EFT operators 
compatible with the specified symmetry 
and comparing their effects to experimental data. 
This yields bounds that restrict the viability of members of the class, 
thereby providing guidance on the acceptability of underlying models. 
The technique is appropriate and powerful in situations 
where no experimental evidence exists for the effects being sought, 
as is the case here.
For spacetime-symmetry violations,
this EFT approach was developed in Ref.\ \cite{kp95}
to study spontaneous Lorentz violation in string theory
\cite{ksp}.
It was subsequently applied to the SM 
to yield the Standard-Model Extension (SME) 
in Ref.\ \cite{ck97}
and generalized to GR coupled to the SM in Ref.\ \cite{ak04}.

Our focus here is on perturbations to GR coupled to the SM
involving EFT terms that have EDV 
while maintaining local Lorentz invariance (LLI).
The perturbative nature and the EDV imply incompatibility 
between the equations of motion and the Bianchi identities 
of Riemann geometry
\cite{ak04,kl21}, 
so these terms can be attributed to models 
within a class of beyond-Riemann theories.
The LLI-EDV symmetry structure is of particular interest because the
corresponding EFT terms lack the severe phenomenological
complications from additional modes that typically arise in
theories with explicit breaking. 
As a result, 
EFT operators can be constructed explicitly for this class of theories 
and can be constrained using experimental and observational data. 
The latter is the primary goal of the present work.

Details of the framework for the EFT construction
have been presented in Refs.\ \cite{ak04,kl21}
and involve several complications,
notably the appearance of additional physical modes
beyond those arising in GR coupled to the SM.
The role of these additional modes
can be reduced by eliminating terms controlling their propagation,
which removes effects of extra long-range forces.
One set of such modes is the antisymmetric tensor $\ch_{\mu\nu}$
associated with local Lorentz violation.
To avoid these modes, 
we restrict the EFT to preserve LLI,
which insures that $\ch_{\mu\nu}$ contains 
only unphysical gauge degrees of freedom.
The other set of additional modes is the vector $\xi_\mu$
associated with diffeomorphism violation.
To incorporate beyond-Riemann effects when LLI holds,
EDV must be present.
The resulting $\xi_\mu$ modes have physical effects,
but their free propagation can be avoided 
by taking the pure-gravity sector to be conventional
\cite{rb15}.
The $\xi_\mu$ modes can then be viewed as nonpropagating auxiliary fields
with derivative couplings in the matter-gravity sector of the EFT. 
To simplify the analysis here,
we assume dynamical torsion and nonmetricity are absent.
However,
the present EFT framework applies also 
to phenomenological and experimental studies 
of fermion couplings involving explicit background torsion 
\cite{torsion}
and nonmetricity
\cite{nonmetricity}.

The laboratory experiments and astrophysical observations
considered here 
involve comparatively weak gravitational fields,
so the linearized limit is sufficient
for a phenomenological analysis of dominant effects from the EFT.
In the linearized EFT,
the local Lorentz and diffeomorphism transformations
combine to yield Lorentz, gauge, and translation transformations
acting in approximately Minkowski spacetime
\cite{kl21}.
For simplicity,
we can limit attention to linearized terms 
that maintain translation invariance (TI).
All such terms exhibit either Lorentz invariance (LI)
or Lorentz violation (LV),
and either gauge invariance (GI)  
or gauge violation (GV).

We remark in passing that the above choices for the EFT
can be matched to the classification 
presented in Table IV of Ref.\ \cite{kl21}.
In this language,
the present work limits attention to EFT terms 
lying both in the row labeled LLI, EDV
and in the columns labeled LI-GI-TI, LV-GV-TI, LI-GV-TI, and LV-GI-TI.
These terms all generically violate the no-go constraints
because they are perturbatively incompatible with the Bianchi identities
of Riemann geometry.
In the context of Fig.\ 2 of Ref.\ \cite{kl21},
the EFT we consider lies in the lower left triangle of the hexagon,
labeled LLI, EDV.

\renewcommand\arraystretch{1.6}
\begin{table*}
\caption{
\label{tab:linferm1}
Terms containing operators of mass dimension $d\leq5$ 
in the linearized fermion Lagrange density $\cL_\ps^\L$.}
\setlength{\tabcolsep}{8pt}
\begin{tabular}{cl}
\hline
\hline
Component & Expression \\
\hline
$	\cL^{\L}_{\ps, 0}	$	&		$		\half(\psb\ga^\mu i\prt_\mu\ps - m\psb\ps)	+	\hc	$	\\
$	\cL^{\L}_{\ps, h}	$	&		$		\quar h\psb\ga^\mu i\prt_\mu\ps - \quar h^{\ka\mu}\psb\ga_\ka i\prt_\mu\ps -\quar mh \psb\ps				
						+	\eigh \ep^{\ka\mu\nu\rh}(\prt_\mu h_{\nu\rh})\psb\ga_5\ga_\ka\ps	+	\hc	$	\\
$	\cL^{(3)\L}_\ps	$	&		$	-	(m^{\prime\L})^{\mu\nu} h_{\mu\nu} \psb \ps 				
						-	i (m^\L_5)^{\mu\nu} h_{\mu\nu} \psb \ga_5\ps 				
						-	(a^\L)^{\ka\mu\nu} h_{\mu\nu}\psb\ga_\ka\ps				
						-	(b^\L)^{\ka\mu\nu}h_{\mu\nu}\psb\gamma_5\ga_\ka\ps				
						-	\half (H^\L)^{\ka\la\mu\nu}h_{\mu\nu}\psb\si_{\ka\la} \ps			$	\\
$	\cL^{(4)\L}_{\ps h}	$	&		$	-	\half (c_h^\L)^{\ka\mu\nu\rh} h_{\nu\rh} \psb \ga_\ka i\prt_\mu \ps				
						-	\half (d_h^\L)^{\ka\mu\nu\rh} h_{\nu\rh} \psb\ga_5\ga_\ka i\prt_\mu \ps			$	\\
			&	\hskip20pt	$	-	\half (e_h^\L)^{\mu\nu\rh}h_{\nu\rh} \psb i\prt_\mu \ps				
						-	\half i (f_h^\L)^{\mu\nu\rh} h_{\nu\rh} \psb \ga_5 i\prt_\mu \ps				
						-	\quar (g_h^\L)^{\ka\la\mu\nu\rh} h_{\nu\rh} \psb \si_{\ka\la} i\prt_\mu \ps	+	\hc	$	\\
$	\cL^{(4)\L}_{\ps \prt h}	$	&		$	-	(c_{\prt h}^\L)^{\ka\mu\nu\rh} (\prt_\mu h_{\nu\rh}) \psb \ga_\ka \ps				
						-	(d_{\prt h}^\L)^{\ka\mu\nu\rh} (\prt_\mu h_{\nu\rh}) \psb \ga_5\ga_\ka \ps			$	\\
			&	\hskip20pt	$	-	(e_{\prt h}^\L)^{\mu\nu\rh} (\prt_\mu h_{\nu\rh}) \psb \ps				
						-	i(f_{\prt h}^\L)^{\mu\nu\rh} (\prt_\mu h_{\nu\rh}) \psb \ga_5\ps				
						-	\half (g_{\prt h}^\L)^{\ka\la\mu\nu\rh} (\prt_\mu h_{\nu\rh}) \psb \si_{\ka\la} \ps			$	\\
$	\cL^{(5)\L}_{\ps h}	$	&		$	-	\half (m_h^{(5)\L})^{\mu\nu\rh\si} h_{\rh\si} \psb i\prt_\mu i\prt_\nu \ps				
						-	\half i(m_{5h}^{(5)\L})^{\mu\nu\rh\si} h_{\rh\si} \psb \ga_5 i\prt_\mu i\prt_\nu \ps			$	\\
			&	\hskip20pt	$	-	\half (a_h^{(5)\L})^{\ka\mu\nu\rh\si} h_{\rh\si} \psb \ga_\ka i\prt_\mu i\prt_\nu \ps				
						-	\half (b_{h}^{(5)\L})^{\ka\mu\nu\rh\si} h_{\rh\si} \psb \ga_5\ga_\ka i\prt_\mu i\prt_\nu \ps			$	\\
			&	\hskip20pt	$	-	\quar (H_{h}^{(5)\L})^{\ka\la\mu\nu\rh\si} h_{\rh\si} \psb \si_{\ka\la} i\prt_\mu i\prt_\nu \ps	+	\hc	$	\\
$	\cL^{(5)\L}_{\ps\prt h}	$	&		$	-	\half (m_{\prt h}^{(5)\L})^{\mu\nu\rh\si} (\prt_\nu h_{\rh\si}) \psb i\prt_\mu \ps				
						-	\half i(m_{5\prt h}^{(5)\L})^{\mu\nu\rh\si} (\prt_\nu h_{\rh\si}) \psb \ga_5 i\prt_\mu \ps			$	\\
			&	\hskip20pt	$	-	\half (a_{\prt h}^{(5)\L})^{\ka\mu\nu\rh\si} (\prt_\nu h_{\rh\si}) \psb \ga_\ka i\prt_\mu \ps				
						-	\half (b_{\prt h}^{(5)\L})^{\ka\mu\nu\rh\si} (\prt_\nu h_{\rh\si}) \psb \ga_5\ga_\ka i\prt_\mu \ps			$	\\
			&	\hskip20pt	$	-	\quar (H_{\prt h}^{(5)\L})^{\ka\la\mu\nu\rh\si} (\prt_\nu h_{\rh\si}) \psb \si_{\ka\la} i\prt_\mu \ps	+	\hc	$	\\
$	\cL^{(5)\L}_{\ps\prt\prt h}	$	&		$	-	(m_{\prt\prt h}^{(5)\L})^{\mu\nu\rh\si} (\prt_\mu\prt_\nu h_{\rh\si}) \psb \ps				
						-	i(m_{5\prt\prt h}^{(5)\L})^{\mu\nu\rh\si} (\prt_\mu\prt_\nu h_{\rh\si}) \psb \ga_5 \ps			$	\\
			&	\hskip20pt	$	-	(a_{\prt\prt h}^{(5)\L})^{\ka\mu\nu\rh\si} (\prt_\mu\prt_\nu h_{\rh\si}) \psb \ga_\ka \ps				
						-	(b_{\prt\prt h}^{(5)\L})^{\ka\mu\nu\rh\si} (\prt_\mu\prt_\nu h_{\rh\si}) \psb \ga_5\ga_\ka \ps			$	\\
			&	\hskip20pt	$	-	\half (H_{\prt\prt h}^{(5)\L})^{\ka\la\mu\nu\rh\si} (\prt_\mu\prt_\nu h_{\rh\si}) \psb \si_{\ka\la} \ps			$	\\
\hline
\hline
\end{tabular}
\end{table*}

\renewcommand\arraystretch{1.6}
\begin{table*}
\caption{
\label{tab:linferm2}
Relationships between coefficients in $\cL_\ps^\L$ and in $\cL_\ps$.}
\setlength{\tabcolsep}{7pt}
\begin{tabular}{ll}
\hline
\hline
$\cL_\ps^\L$ & $\cL_\ps$ \\
\hline
$	(m^{\prime \L})^{\mu\nu}	$	&		$		(\ov{m}^{\prime \L})^{\mu\nu}		
						+	\half \ov{m}\pr_\eff \et^{\mu\nu}	$	\\
$	(m_5^\L)^{\mu\nu}	$	&		$		(\ov{m}_5^{\L})^{\mu\nu}		
						+	\half \ov{m}_{5\eff} \et^{\mu\nu}	$	\\
$	(a^\L)^{\ka\mu\nu}	$	&		$		(\ov{a}^{\L})^{\ka\mu\nu}		
						+	\half \ov{a}^\ka_\eff \et^{\mu\nu}		
						+	\quar (\ov{a}_\eff^{\mu} \et^{\nu\ka}		
						+	\ov{a}_\eff^{\nu} \et^{\mu\ka})	$	\\
$	(b^\L)^{\ka\mu\nu}	$	&		$		(\ov{b}^{\L})^{\ka\mu\nu}		
						+	\half \ov{b}^\ka_\eff \et^{\mu\nu}		
						+	\quar (\ov{b}_\eff^{\mu} \et^{\nu\ka}		
						+	\ov{b}_\eff^{\nu} \et^{\mu\ka})	$	\\
$	(H^\L)^{\ka\la\mu\nu}	$	&		$		(\ov{H}^{\L})^{\ka\la\mu\nu}		
						+	\half \ov{H}^{\ka\la} _\eff\et^{\mu\nu}		
						+	\quar [(\ov{H}_\eff^{\mu\la} \et^{\ka\nu}		
						+	\ov{H}_\eff^{\nu\la} \et^{\ka\mu})		
						-	(\ka\leftrightarrow\la)]	$	\\
$	(c_h^\L)^{\ka\mu\nu\rh}	$	&		$		(\ov{c}^{\L})^{\ka\mu\nu\rh}		
						+	\half \ov{c}_\eff^{\ka\mu} \et^{\nu\rh}		
						+	\quar(\ov{c}_\eff^{\nu\mu}\et^{\rh\ka}		
						+	\ov{c}_\eff^{\rh\mu}\et^{\nu\ka})	$	\\
$	(d_h^\L)^{\ka\mu\nu\rh}	$	&		$		(\ov{d}^{ \L})^{\ka\mu\nu\rh}		
						+	\half \ov{d}_\eff^{\ka\mu} \et^{\nu\rh}		
						+	\quar(\ov{d}_\eff^{\nu\mu}\et^{\rh\ka}		
						+	\ov{d}_\eff^{\rh\mu}\et^{\nu\ka})	$	\\
$	(e_h^\L)^{\mu\nu\rh}	$	&		$		(\ov{e}^{ \L})^{\mu\nu\rh}		
						+	\half \ov{e}_\eff^{\mu} \et^{\nu\rh}	$	\\
$	(f_h^\L)^{\mu\nu\rh}	$	&		$		(\ov{f}^{ \L})^{\mu\nu\rh}		
						+	\half \ov{f}_\eff^{\mu} \et^{\nu\rh}	$	\\
$	(g_h^\L)^{\ka\la\mu\nu\rh}	$	&		$		(\ov{g}^{ \L})^{\ka\la\mu\nu\rh}		
						+	\half \ov{g}_\eff^{\ka\la\mu} \et^{\nu\rh}		
						+	\quar [(\ov{g}_\eff^{\nu\la\mu} \et^{\ka\rh}		
						+	\ov{g}_\eff^{\rh\la\mu} \et^{\ka\nu})		
						-	(\ka\leftrightarrow\la)]	$	\\
$	(c_{\prt h}^\L)^{\ka\mu\nu\rh}	$	&		$		\eigh (\ov{d}_\eff^{\al\nu}\et_{\al\be}\ep^{\be\mu\rh\ka}		
						+	\ov{d}_\eff^{\al\rh}\et_{\al\be}\ep^{\be\mu\nu\ka})	$	\\
$	(d_{\prt h}^\L)^{\ka\mu\nu\rh}	$	&		$		\eigh (\ov{c}_\eff^{\al\nu}\et_{\al\be}\ep^{\be\mu\rh\ka}		
						+	\ov{c}_\eff^{\al\rh}\et_{\al\be}\ep^{\be\mu\nu\ka})	$	\\
$	(e_{\prt h}^\L)^{\mu\nu\rh}	$	&		$		\quar (\ov{g}_\eff^{\mu\nu\rh}		
						+	\ov{g}_\eff^{\mu\rh\nu})	$	\\
$	(f_{\prt h}^\L)^{\mu\nu\rh}	$	&		$	-	\eigh (\ov{g}_\eff^{\al\be\nu} \et_{\al\ga} \et_{\be\de} \ep^{\ga\de\mu\rh}		
						+	\ov{g}_\eff^{\al\be\rh} \et_{\al\ga} \et_{\be\de} \ep^{\ga\de\mu\nu})	$	\\
$	(g_{\prt h}^\L)^{\ka\la\mu\nu\rh}	$	&		$		\eigh [(\ov{e}_\eff^\nu \et^{\ka\mu} \et^{\la\rh}		
						+	\ov{e}_\eff^\rh \et^{\ka\mu} \et^{\la\nu})		
						-	(\ka\leftrightarrow\la)]		
						+	\eigh (\ov{f}_\eff^\nu \ep^{\ka\la\mu\rh}		
						+	\ov{f}_\eff^\rh \ep^{\ka\la\mu\nu})	$	\\
$	(m_h^{(5)\L})^{\mu\nu\rh\si}	$	&		$		(\ov{m}^{(5)\L})^{\mu\nu\rh\si}		
						+	\half (\ov{m}_\eff^{(5)})^{\mu\nu} \et^{\rh\si}	$	\\
$	(m_{5h}^{(5)\L})^{\mu\nu\rh\si}	$	&		$		(\ov{m}^{(5)\L}_5)^{\mu\nu\rh\si}		
						+	\half (\ov{m}_{5\eff}^{(5)})^{\mu\nu} \et^{\rh\si}	$	\\
$	(a_{h}^{(5)\L})^{\ka\mu\nu\rh\si}	$	&		$	-	(\ov{a}^{(5)\L})^{\ka\mu\nu\rh\si}		
						-	\half (\ov{a}_\eff^{(5)})^{\ka\mu\nu} \et^{\rh\si}		
						-	\quar[(\ov{a}_\eff^{(5)})^{\rh\mu\nu} \et^{\ka\si}		
						+	(\ov{a}_\eff^{(5)})^{\si\mu\nu} \et^{\ka\rh}]	$	\\
$	(b_{h}^{(5)\L})^{\ka\mu\nu\rh\si}	$	&		$	-	(\ov{b}^{(5)\L})^{\ka\mu\nu\rh\si}		
						-	\half (\ov{b}_\eff^{(5)})^{\ka\mu\nu} \et^{\rh\si}		
						-	\quar[(\ov{b}_\eff^{(5)})^{\rh\mu\nu} \et^{\ka\si}		
						+	(\ov{b}_\eff^{(5)})^{\si\mu\nu} \et^{\ka\rh}]	$	\\
$	(H_{h}^{(5)\L})^{\ka\la\mu\nu\rh\si}	$	&		$		(\ov{H}^{(5)\L})^{\ka\la\mu\nu\rh\si}		
						+	\half (\ov{H}_\eff^{(5)})^{\ka\la\mu\nu} \et^{\rh\si}		
						+	\quar\big[[(\ov{H}_\eff^{(5)})^{\rh\la\mu\nu} \et^{\ka\si}		
						+	(\ov{H}_\eff^{(5)})^{\si\la\mu\nu} \et^{\ka\rh}]		
						-	(\ka\leftrightarrow\la)\big]	$	\\
$	(m_{\prt h}^{(5)\L})^{\mu\nu\rh\si}	$	&		$		\half [(\ov{H}_\eff^{(5)})^{\nu\si\mu\rh}		
						+	(\ov{H}_\eff^{(5)})^{\nu\rh\mu\si}]	$	\\
$	(m_{5\prt h}^{(5)\L})^{\mu\nu\rh\si}	$	&		$	-	\quar [(\ov{H}_\eff^{(5)})^{\al\be\mu\rh} \et_{\al\ga} \et_{\be\de} \ep^{\ga\de\nu\si}		
						+	(\ov{H}_\eff^{(5)})^{\al\be\mu\si} \et_{\al\ga} \et_{\be\de} \ep^{\ga\de\nu\rh}]	$	\\
$	(a_{\prt h}^{(5)\L})^{\ka\mu\nu\rh\si}	$	&		$		\quar [(\ov{b}_\eff^{(5)})^{\al\mu\rh} \et_{\al\be} \ep^{\be\nu\si\ka}		
						+	(\ov{b}_\eff^{(5)})^{\al\mu\si} \et_{\al\be} \ep^{\be\nu\rh\ka}]	$	\\
$	(b_{\prt h}^{(5)\L})^{\ka\mu\nu\rh\si}	$	&		$		\quar [(\ov{a}_\eff^{(5)})^{\al\mu\rh} \et_{\al\be} \ep^{\be\nu\si\ka}		
						+	(\ov{a}_\eff^{(5)})^{\al\mu\si} \et_{\al\be} \ep^{\be\nu\rh\ka}]	$	\\
$	(H_{\prt h}^{(5)\L})^{\ka\la\mu\nu\rh\si}	$	&		$		\quar \big[[(\ov{m}_\eff^{(5)})^{\mu\rh} \et^{\nu\ka} \et{\si\la}		
						+	(\ov{m}_\eff^{(5)})^{\mu\si} \et^{\nu\ka} \et^{\rh\la}]		
						-	(\ka\leftrightarrow\la)\big]		
						+	\quar [(\ov{m_5}_\eff^{(5)})^{\mu\rh} \ep^{\ka\la\nu\si}		
						+	(\ov{m_5}^{(5)})^{\mu\si} \ep^{\ka\la\nu\rh}]	$	\\
$	(m_{\prt\prt h}^{(5)\L})^{\mu\nu\rh\si}	$	&		$		\half [(\ov{m}_{R,\eff}^{(5)})^{\mu\rh\si\nu}		
						+	(\ov{m}_{R,\eff}^{(5)})^{\nu\rh\si\mu}]		
						+	\quar (\ov{m}_\eff^{(5)})^{\rh\si} \et^{\mu\nu}		
						-	\eigh \big[[(\ov{m}_\eff^{(5)})^{\mu\rh} \et^{\nu\si}		
						+	(\ov{m}_\eff^{(5)})^{\mu\si} \et^{\nu\rh}]		
						+	(\mu\leftrightarrow\nu)\big]	$	\\
$	(m_{5\prt\prt h}^{(5)\L})^{\mu\nu\rh\si}	$	&		$		\half [(\ov{m}_{5R,\eff}^{(5)})^{\mu\rh\si\nu}		
						+	(\ov{m}_{5R,\eff}^{(5)})^{\nu\rh\si\mu}]		
						+	\quar (\ov{m}_{5\eff}^{(5)})^{\rh\si} \et^{\mu\nu}		
						-	\eigh \big[[(\ov{m}_{5\eff}^{(5)})^{\mu\rh} \et^{\nu\si}		
						+	(\ov{m}_{5\eff}^{(5)})^{\mu\si} \et^{\nu\rh}]		
						+	(\mu\leftrightarrow\nu)\big]	$	\\
$	(a_{\prt\prt h}^{(5)\L})^{\ka\mu\nu\rh\si}	$	&		$	-	\half [(\ov{a}_{R,\eff}^{(5)})^{\ka\mu\rh\si\nu}		
						+	(\ov{a}_{R,\eff}^{(5)})^{\ka\nu\rh\si\mu}]		
						-	\quar (\ov{a}_\eff^{(5)})^{\ka\rh\si} \et^{\mu\nu}		
						+	\eigh \big[[(\ov{a}_\eff^{(5)})^{\ka\mu\rh} \et^{\nu\si}		
						+	(\ov{a}_\eff^{(5)})^{\ka\mu\si} \et^{\nu\rh}]		
						+	(\mu\leftrightarrow\nu)\big]	$	\\
$	(b_{\prt\prt h}^{(5)\L})^{\ka\mu\nu\rh\si}	$	&		$	-	\half [(\ov{b}_{R,\eff}^{(5)})^{\ka\mu\rh\si\nu}		
						+	(\ov{b}_{R,\eff}^{(5)})^{\ka\nu\rh\si\mu}]		
						-	\quar (\ov{b}_\eff^{(5)})^{\ka\rh\si} \et^{\mu\nu}		
						+	\eigh \big[[(\ov{b}_\eff^{(5)})^{\ka\mu\rh} \et^{\nu\si}		
						+	(\ov{b}_\eff^{(5)})^{\ka\mu\si} \et^{\nu\rh}]		
						+	(\mu\leftrightarrow\nu)\big]	$	\\
$	(H_{\prt\prt h}^{(5)\L})^{\ka\la\mu\nu\rh\si}	$	&		$		\half [(\ov{H}_{R,\eff}^{(5)})^{\ka\la\mu\rh\si\nu}		
						+	(\ov{H}_{R,\eff}^{(5)})^{\ka\la\nu\rh\si\mu}]		
						+	\quar (\ov{H}_\eff^{(5)})^{\ka\la\rh\si} \et^{\mu\nu}		
						-	\eigh \big[[(\ov{H}_\eff^{(5)})^{\ka\la\mu\rh} \et^{\nu\si}		
						+	(\ov{H}_\eff^{(5)})^{\ka\la\mu\si} \et^{\nu\rh}]		
						+	(\mu\leftrightarrow\nu)\big]	$	\\
\hline
\hline
\end{tabular}
\end{table*}

For our phenomenological analyses,
we focus on leading-order effects arising from the propagation 
of a Dirac fermion $\ps$ of mass $m$ in the presence
of a weak gravitational field 
with metric $g_{\mu\nu} = \et_{\mu\nu} + h_{\mu\nu}$.
The metric fluctuation $h_{\mu\nu}$ 
includes contributions from the derivatives $\prt_\la \xi_\mu$ 
of the $\xi_\mu$ modes.
Note, however,
that in the EFT context these contributions 
are determined by the backgrounds $k$
and represent small corrections to the GR value of $h_{\mu\nu}$.
The weak-field assumption implies that the latter is already small,
so for many applications 
$h_{\mu\nu}$ can be approximated at leading order by its GR value.
For simplicity,
we neglect possible couplings to derivatives $Dk$ 
of the backgrounds $k$. 
It then suffices to consider quadratic fermion terms
and their gravitational couplings,
allowing for arbitrary LLI-EDV operators.
All the relevant fermion-gravity terms involving operators with $d\leq 6$ 
and without background derivatives
are presented in Table XI of Ref.\ \cite{kl21}.
As $d$ increases,
these terms acquire more fermion derivatives
and hence can be expected to generate suppressed effects
in laboratory experiments because the relevant fermion momenta 
are small compared to the Planck scale.
Nonetheless,
to capture effects from both minimal and nonminimal terms,
the analysis that follows includes terms containing operators with $d\leq 5$.

Implementing the linearization to extract all fermion terms 
with operators of mass dimension $d\leq 5$
in the Lagrange density $\cL_\ps^\L$ of the linearized EFT,
we find the results displayed in Table \ref{tab:linferm1}.
In the table,
$\cL_\ps^\L$ is separated into pieces
containing terms with specified $d$ and number of derivatives of $h_{\mu\nu}$.
These pieces are listed in the first column,
while the second column shows the explicit form
of the corresponding terms. 
The first two rows in the table represent the linearization
of the usual Lagrange density for a massive Dirac fermion coupled to gravity,
and they lie in the LI-GI-TI class.
All other terms in the table exhibit LV, GV or both.
The only LI terms are ones with coefficients
constructed from the Minkowski metric $\et_{\mu\nu}$ 
and the Levi-Civita tensor $\ep_{\ka\la\mu\nu}$.
The only GI terms involve combinations of two derivatives $\prt\prt h$ 
of the metric fluctuation that arise from the curvature tensor. 
The notation for each coefficient appearing in the table
is chosen in accordance with standard usage in the literature,
with a superscript $\L$ indicating 
that the corresponding operator is linearized.
The primary letter on a coefficient distinguishes the spin 
and charge-conjugation, parity-inversion, and time-reversal (CPT) properties
of the dynamical operator,
while the subscript indicates the number of derivatives 
of $h_{\mu\nu}$ it contains.

The terms listed in Table \ref{tab:linferm1} are obtained
by linearization of the full fermion-gravity Lagrange density $\cL_\ps$ 
provided in Table XI of Ref.\ \cite{kl21}.
The coefficients appearing in Table \ref{tab:linferm1}
are therefore combinations of relevant parts of the breve coefficients 
appearing in Table XI of Ref.\ \cite{kl21}.
Each breve coefficient is a linear combination of backgrounds 
contracted with vierbeins, metrics, and Levi-Civita tensors. 
In the linearized scenario relevant to the EFT of interest here,
a generic breve coefficient $\breve{k}^{\cdots}$
can be written as a sum of two parts involving explicit backgrounds,
\bea
\breve{k}^{\cdots} 
&\equiv& \ol{k}^{\cdots} + \ol{k}^{\cdots\mu\nu} g_{\mu\nu} + \ldots
\nn\\
& \approx &
\ol{k}_\eff^{\cdots} + (\ol{k}^\L)^{\cdots\mu\nu} h_{\mu\nu},
\label{asy}
\eea
where the background $\ol{k}_\eff^{\cdots}$
appearing in the approximately Minkowski spacetime is
the breve coefficient $\breve{k}^{\cdots}$
taken at zeroth order in vierbein and metric fluctuations,
\bea
\ol{k}_\eff^{\cdots} &\equiv& 
\ol{k}^{\cdots} + \ol{k}^{\cdots\mu\nu} \et_{\mu\nu} + \ldots,
\nn\\
(\ol{k}^\L)^{\cdots\mu\nu} &\equiv& \ol{k}^{\cdots\mu\nu} + \ldots.
\eea
For example,
the breve coefficient $\breve{a}^\ka$ appearing 
in the piece $\cL_\ps^{(3)}$ of $\cL_\ps$
reduces in the present EFT context to 
$\breve{a}^\ka =\ov{a}^\ka_\eff+(\ol{a}^\L)^{\ka\mu\nu}h_{\mu\nu}$,
with 
$\ov{a}^\ka_\eff\equiv \ov{a}^\ka+\ov{a}^{\ka\mu\nu}\et_{\mu\nu}+\ldots$
and $(\ol{a}^\L)^{\ka\mu\nu} \equiv \ol{a}^{\ka\mu\nu} + \ldots$.
Note that our assumption of TI for the linearized theory
implies that all coefficients considered here are spacetime constants.

The explicit relationships between 
the linearized coefficients appearing in Table \ref{tab:linferm1} 
and the breve coefficients appearing in $\cL_\ps$
are provided in Table \ref{tab:linferm2}.
The first column of this table displays the linearized coefficients
appearing in $\cL_\ps^\L$,
while the second column establishes 
the link to the explicit backgrounds 
contained in the breve coefficients appearing in $\cL_\ps$
and defined via Eq.\ \rf{asy}.
Note that the asymptotic parts of certain breve coefficients are absent
in Table \ref{tab:linferm2}
because they contribute to the linearized EFT only for $d\geq 6$.

\subsection{Nonrelativistic hamiltonian}
\label{Nonrelativistic hamiltonian}

The linearized Lagrange density $\cL_\ps^\L$ 
given in Table \ref{tab:linferm1} 
can be used as the basis for phenomenological analyses.
However,
many laboratory experiments sensitive to fermion-gravity couplings
involve slow-moving particle species 
experiencing the gravitational field of the Earth.
For these types of experiments,
the analysis of data for signals of physics beyond Riemann gravity
involves the nonrelativistic particle hamiltonian $H$.
This can be extracted from the linearized Lagrange density $\cL_\ps^\L$ 
via a generalized Foldy-Wouthuysen transformation
\cite{fw50},
using techniques established for backgrounds 
violating spacetime symmetries
\cite{kl99jmp,kl01,gos09,kt11,km12,km13,yb13,cl16,zx18}.

At leading order in the backgrounds,
the perturbative relativistic hamiltonian
can be identified from the Euler-Lagrange equations
obtained by variation of the linearized action
\cite{km12}.
In approximately flat spacetime,
this bypasses the complications of nonstandard time evolution
introduced by certain background components
and hence avoids the necessity for prior field redefinitions
\cite{kt11}
or modifications to the inner product in the Hilbert space
\cite{lp80}. 
The relativistic hamiltonian can then be block diagonalized
at the desired order in the particle 3-momentum $p_j = -i\prt_j$
using an iterative method,
and the nonrelativistic hamiltonian can be extracted 
from the upper $2\times2$ block 
\cite{kl99jmp}.
The results obtained via this procedure
generalize those in Ref.\ \cite{km13}
obtained for a Dirac fermion in Minkowski spacetime
in the presence of Lorentz-violating operators of arbitrary mass dimension $d$.

With the above techniques,
the derivation of the nonrelativistic hamiltonian $H$
from the linearized Lagrange density $\cL^\L_\ps$ in Table \ref{tab:linferm1} 
is lengthy but straightforward.
It is convenient to express the result as a sum of pieces,
\beq
H=H_0 + H_{\ph} + H_{\si\ph} + H_{g} + H_{\si g} + \ldots,
\label{hamiltonian}
\eeq
where $H_0$ is the hamiltonian in the absence of backgrounds.
In this sum,
the spin-dependent terms containing the Pauli spin matrices $\vec \si$
are identified with a subscript $\si$.
The perturbative corrections of this type represent 
anomalous spin-gravity couplings
and in this context can be viewed as WEP violations.
The pieces with a subscript $\ph$
depend directly on the gravitational potential $\ph \approx -h_{00}/2$,
while those with a subscript $g$
depend only on the gravitational acceleration 
$\vec g \equiv -\vec \nabla \ph$.
The ellipsis indicates terms that depend on higher derivatives of $\ph$.

For applications to laboratory experiments,
it typically suffices to take the gravitational acceleration in the laboratory 
as uniform and directed along $-\hat{z}$ in the canonical laboratory frame,
$\vec g = - g \hat z$,
so the gravitational potential is $\ph=-\vec g \cdot \vec z = gz$.
We incorporate here relativistic corrections to second order in $p_j$.
With these approximations,
we can extract explicit forms for the various terms 
in the hamiltonian \rf{hamiltonian}.

For the piece $H_0$,
the procedure generates the expression 
\beq
H_0 = \frac{\vec{p}^{\,2}}{2m}-m\vec{g}\cdot\vec{z}
-\frac{3}{4m}(\vec{p}^{\,2}\vec{g}\cdot\vec{z}
+\vec{g}\cdot\vec{z}~\vec{p}^{\,2})
+ \frac{3}{4m} (\vec{\si}\times\vec{p})\cdot\vec{g}.
\label{hzero}
\eeq
The first two terms on the right-hand side
are the usual strict nonrelativistic limit.
The third term is the leading-order relativistic correction,
while the last term is the spin-orbit coupling.
Note that no Darwin-type term 
proportional to the divergence of $\vec g$
appears because $\vec g$ is uniform by assumption.
The form of $H_0$ has been the subject of numerous investigations 
in the literature 
\cite{ot62,as78,ffc81,hn90,cp91,jh94,vr00,yo01,st05}.
Our result \rf{hzero} matches Eq.\ (14) in Ref.\ \cite{st05},
which was derived for uniform acceleration
and expressed in the physical Foldy-Wouthuysen representation.

The spin-independent piece of $H$ 
coupling via the gravitational potential $\ph$
can be written in the form
\bea
H_{\ph}&=&(k^\NR_{\ph}) \vec{g}\cdot\vec{z}
+(k^\NR_{\ph p})^j \half(p^j \vec{g}\cdot\vec{z}
+\vec{g}\cdot\vec{z}\ p^j)
\nn\\
&&
+(k^\NR_{\ph pp})^{jk} \half(p^j p^k \vec{g}\cdot\vec{z}
+\vec{g}\cdot\vec{z}\ p^j p^k),
\label{eq:ph}
\eea
where the coefficient $(k^\NR_{\ph pp})^{jk}$ is defined to be symmetric,
\beq
(k^\NR_{\ph pp})^{jk}=(k^\NR_{\ph pp})^{kj}.
\eeq
Each term in Eq.\ \rf{eq:ph} depends on position $\vec z$
through the dependence on the gravitational potential.
The coefficients 
$(k^\NR_{\ph})$, $(k^\NR_{\ph p})^j$, and $(k^\NR_{\ph pp})^{jk}$
are spacetime constants 
that control the magnitude of the effects
produced by the operators in $H_\ph$.
The superscripts $\NR$ serve as a reminder
that the coefficients are defined in the nonrelativistic limit,
while the subscripts $\ph$ and $p$ reflect 
the dependence on the potential and on the fermion momentum.
The term with coefficient $(k^\NR_{\ph})$
and the component of the third term 
governed by the trace $(k^\NR_{\ph pp})_j{}^j$ 
of $(k^\NR_{\ph pp})^{jk}$
are invariant under rotations,
and they represent EFT contributions
that shift the sizes of the second and third terms in $H_0$. 
The remaining terms in $H_\ph$ violate rotation symmetry.
The term in $H_\ph$ with coefficient $(k^\NR_{\ph p})^j$ 
violates parity P and time reversal T,
while the other two are P and T even.

The spin-dependent piece of $H$ coupling via the gravitational potential $\ph$
is given by
\bea
H_{\si\ph}&=&(k^{\NR}_{\si\ph})^j \si^j \vec{g}\cdot\vec{z}
+(k^{\NR}_{\si\ph p})^{jk} \half \si^j (p^k \vec{g}\cdot\vec{z}
+\vec{g}\cdot\vec{z}\ p^k)
\nn\\
&&
+(k^{\NR}_{\si\ph pp})^{jkl} \half \si^j (p^k p^l \vec{g}\cdot\vec{z}
+\vec{g}\cdot\vec{z}\ p^k p^l).
\label{eq:siph}
\eea
In this equation,
the coefficient $(k^{\NR}_{\si\ph pp})^{jkl}$
is defined to be symmetric on the last two indices,
\beq
(k^{\NR}_{\si\ph pp})^{jkl}=(k^{\NR}_{\si\ph pp})^{jlk},
\label{indexid}
\eeq
and all the coefficients are spacetime constants.
Each term in $H_{\si\ph}$ inherits dependence on the position $\vec z$
from the gravitational potential.
The second term contains a rotation-invariant component 
controlled by the trace $(k^{\NR}_{\si\ph p})_j{}^j$.
The corresponding operator $(\vec \si \cdot \vec p)(\vec g \cdot \vec z)$ 
represents rotation-invariant effects 
that are distinct from those appearing in the hamiltonian $H_0$
with vanishing backgrounds.
The component of $(k^{\NR}_{\si\ph pp})^{jkl}$ proportional to $\ep^{jkl}$
is absent in $H_{\si\ph}$ 
because the corresponding operator vanishes identically.
The first and third terms in $H_{\si\ph}$ are P even and T odd,
while the second term is P odd and T even.

Turning next to the pieces of $H$ coupling 
via the gravitational acceleration $\vec g$,
we find that the spin-independent piece $H_{g}$ 
can be written in the form
\beq
H_{g}=(k^\NR_{g})^j g^j+(k^\NR_{gp})^{jk} p^j g^k 
+(k^\NR_{gpp})^{jkl}p^j p^k g^l.
\label{eq:g}
\eeq
where we define 
\beq
(k^\NR_{gpp})^{jkl} = (k^\NR_{gpp})^{kjl}.
\eeq
All three terms in the expression \rf{eq:g} are position independent.
All are rotation violating,
except for the operator $\vec p \cdot \vec g$ in the second term 
associated with the trace $(k^\NR_{gp})_j{}^j$.
Note that in principle a totally antisymmetric component 
of $(k^\NR_{gpp})^{jkl}\propto \ep^{jkl}$ 
would govern rotation-invariant effects,
but the corresponding operator $(\vec p \times \vec p) \cdot \vec g$ 
vanishes identically.
The second term in $H_{g}$ is P even and T odd,
while the other two are P odd and T even. 

\renewcommand\arraystretch{1.6}
\begin{table*}
\caption{
\label{tab:Hamiltonian}
Correspondence between nonrelativistic and linearized coefficients.}
\setlength{\tabcolsep}{3pt}
\begin{tabular}{ll}
\hline
\hline
NR coefficient & Linearized coefficient \\
\hline
$	(k^\NR_{\ph})	$	&		$		2(m^{\prime\L})^{\aaa}-2(a^\L)^{t\aaa}			
						+	2m(e^\L_h)^{t\aaa}-2m(c^\L_h)^{tt\aaa}			
						+	2m^2(m^{(5)\L}_h)^{tt\aaa}-2m^2(a^{(5)\L}_h)^{ttt\aaa}	$	\\	
$	(k^\NR_{\ph p})^j	$	&		$		\tfrac{2}{m} (a^\L)^{j\aaa}			
						-	2(e^\L_h)^{j\aaa}+2(c^\L_h)^{jt\aaa}+2(c^\L_h)^{tj\aaa}			
						-	4m(m^{(5)\L}_h)^{jt\aaa}+2m(a^{(5)\L}_h)^{jtt\aaa}+4m(a^{(5)\L}_h)^{tjt\aaa}	$	\\	
$	(k^\NR_{\ph pp})^{jk}	$	&		$	-	\tfrac1m[(c^\L_h)^{jk\aaa}+(c^\L_h)^{kj\aaa}]+2(m^{(5)\L}_h)^{jk\aaa}			
						-	2(a^{(5)\L}_h)^{tjk\aaa}-2[(a^{(5)\L}_h)^{jkt\aaa}+(a^{(5)\L}_h)^{kjt\aaa}]	$	\\	
			&	\hskip20pt	$	-	\de^{jk}\big[\tfrac{1}{m^2}(m^{\prime\L})^{\aaa}+\tfrac{1}{m} (c_h^\L)^{tt\aaa}			
						-	(m^{(5)\L}_h)^{tt\aaa}+2(a^{(5)\L}_h)^{ttt\aaa}\big]	$	\\	
$	(k^\NR_{\si \ph})^{j}	$	&		$	-	2(b^\L)^{j\aaa}+\ep^{jkl}(H^\L)^{kl\aaa}			
						-	2m(d^\L_h)^{jt\aaa}+m\ep^{jkl}(g^\L_h)^{klt\aaa}			
						-	2m^2(b^{(5)\L}_h)^{jtt\aaa}+m^2\ep^{jkl}(H^{(5)\L}_h)^{kltt\aaa}	$	\\	
$	(k^\NR_{\si \ph p})^{jk}	$	&		$		2(d^\L_h)^{jk\aaa}-\ep^{jmn}(g^\L_h)^{mnk\aaa}			
						+	4m(b^{(5)\L}_h)^{jkt\aaa}-2m\ep^{jmn}(H^{(5)\L}_h)^{mnkt\aaa}	$	\\	
			&	\hskip20pt	$	+	\de^{jk}\big[\tfrac2m(b^\L)^{t\aaa}+2(d^\L_h)^{tt\aaa}+2m(b^{(5)\L}_h)^{ttt\aaa}\big]			
						-	\ep^{jkl}\big[\tfrac2m(H^\L)^{tl\aaa}+2(g^\L_h)^{tlt\aaa}+2m(H^{(5)\L}_h)^{tltt\aaa}\big]	$	\\	
$	(k^\NR_{\si \ph pp})^{jkl}	$	&		$	-	2(b^{(5)\L}_h)^{jkl\aaa}+\ep^{jmn}(H^{(5)\L}_h)^{mnkl\aaa}			
						+	\de^{kl}\big[\tfrac1{m^2}(b^\L)^{j\aaa}+\tfrac1{2m}\ep^{jmn}(g^\L_h)^{mnt\aaa}			
						-	(b^{(5)\L}_h)^{jtt\aaa}+\ep^{jkl}(H^{(5)\L}_h)^{mntt\aaa}\big]	$	\\	
			&	\hskip20pt	$	+	\half\Big[\Big(-\de^{jk}\big[\tfrac1{m^2}(b^\L)^{l\aaa}+\tfrac1{2m^2}\ep^{lmn}(H^\L)^{mn\aaa}			
						+	\tfrac2m(d^L_h)^{tl\aaa}+\tfrac1m(d^\L_h)^{lt\aaa}+\tfrac1{2m}\ep^{lmn}(g^\L_h)^{mnt\aaa}	$	\\	
			&	\hskip40pt	$	+	4(b^{(5)\L}_h)^{ttl\aaa}+(b^{(5)\L}_h)^{ltt\aaa}+\half\ep^{lmn}(H^{(5)\L}_h)^{mntt\aaa}\big]			
						+	\ep^{jkm}\big[\tfrac2m(g^\L_h)^{tml\aaa}+4(H^{(5)\L}_h)^{tmlt\aaa}\big]\Big)			
						+	(k\leftrightarrow l)\Big]	$	\\	
$	(k^\NR_g)^j	$	&		$		\tfrac{1}{m}(H^\L)^{tj\aaa}			
						+	2(e^\L_{\prt h})^{j\aaa}-2(c^\L_{\prt h})^{tj\aaa}+(g^\L_h)^{tjt\aaa}			
						+	2m(m^{(5)\L}_{\prt h})^{tj\aaa}-2m(a^{(5)\L}_{\prt h})^{ttj\aaa}			
						+	m(H^{(5)\L}_h)^{tjtt\aaa}	$	\\	
$	(k^\NR_{gp})^{jk}	$	&		$		\tfrac{2}{m}(c^\L_{\prt h})^{jk\aaa}-\tfrac{1}{m}(g^\L_h)^{tkj\aaa}			
						-	2(m^{(5)\L}_{\prt h})^{jk\aaa}+2(a^{(5)\L}_{\prt h})^{tjk\aaa}			
						+	2(a^{(5)\L}_{\prt h})^{jtk\aaa}-2(H^{(5)\L}_h)^{tkjt\aaa}	$	\\	
			&	\hskip20pt	$	-	\ep^{jkl}[\tfrac{1}{2m^2}(b^\L)^{l\aaa}+\tfrac{1}{2m}(d^\L_h)^{lt\aaa}+\half(b^{(5)\L}_h)^{ltt\aaa}]			
						-	\ep^{jkl}\ep^{lmn}[\tfrac{1}{4m^2}(H^\L)^{mn\aaa}			
						+	\tfrac{1}{4m}(g^\L_h)^{mnt\aaa}+\quar(H^{(5)\L}_h)^{mntt\aaa}]	$	\\	
$	(k^\NR_{gpp})^{jkl}	$	&		$	-	\tfrac1m(a^{(5)\L}_{\prt h})^{jkl\aaa}-\tfrac1m(a^{(5)\L}_{\prt h})^{kjl\aaa}+\tfrac1m(H^{(5)\L}_h)^{tljk\aaa}	$	\\	
			&	\hskip20pt	$	-	\de^{jk}\big[\tfrac1{m^2}(e^\L_{\prt h})^{l\aaa}-\tfrac1{2m^2}(g^\L_h)^{tlt\aaa}			
						+	\tfrac1m(a^{(5)\L}_{\prt h})^{ttl\aaa}-\tfrac1m(H^{(5)\L}_h)^{tltt\aaa}\big]	$	\\	
			&	\hskip20pt	$	+	\ep^{jlm}[\tfrac{1}{2m^2}(d^\L_h)^{mk\aaa}+\tfrac{1}{m}(b^{(5)\L}_h)^{mtk\aaa}]			
						+	\ep^{jlm}\ep^{mnr}[\tfrac{1}{4m^2}(g^\L_h)^{nrk\aaa}+\tfrac{1}{2m}(H^{(5)\L}_h)^{nrtk\aaa}]	$	\\	
$	(k^\NR_{\si g})^{jk}	$	&		$	-	2(d^\L_{\prt h})^{jk\aaa}+\ep^{jmn}(g^\L_{\prt h})^{mnk\aaa}			
						-	2m(b^{(5)\L}_{\prt h})^{jtk\aaa}+m\ep^{jmn}(H^{(5)\L}_{\prt h})^{mntk\aaa}	$	\\	
			&	\hskip20pt	$	-	\de^{jk}\big[\tfrac1m(m^\L_5)^{\aaa}+(f^\L_h)^{t\aaa}+m(m^{(5)\L}_{5h})^{tt\aaa}\big]			
						+	\ep^{jkl}\big[\tfrac1m(a^\L)^{l\aaa}+(c^\L_h)^{lt\aaa}+m(a^{(5)\L}_h)^{ltt\aaa}\big]	$	\\	
$	(k^\NR_{\si gp})^{jkl}	$	&		$		2(b^{(5)\L}_{\prt h})^{jkl\aaa}-\ep^{jmn}(H^{(5)\L}_{\prt h})^{mnkl\aaa}			
						+	\ep^{jkl}\big[\tfrac1{2m^2}(m^{\prime\L})^{\aaa}+\tfrac1{2m^2}(a^\L)^{t\aaa}			
						+	\tfrac1{2m}(e^\L_h)^{t\aaa}+\tfrac1{2m}(c^\L_h)^{tt\aaa}	$	\\	
			&	\hskip20pt	$	+	\half(m^{(5)\L}_h)^{tt\aaa}+\half(a^{(5)\L}_h)^{ttt\aaa}\big]			
						+	\de^{jk}\big[\tfrac2m(d^\L_{\prt h})^{tl\aaa}+2(b^{(5)\L}_{\prt h})^{ttl\aaa}\big]			
						+	\de^{jl}\big[\tfrac1m(f^\L_h)^{k\aaa}+2(m^{(5)\L}_{5h})^{kt\aaa}\big]	$	\\	
			&	\hskip20pt	$	-	\ep^{jkm}\big[\tfrac2m(g^\L_{\prt h})^{tml\aaa}+2(H^{(5)\L}_{\prt h})^{tmtl\aaa}\big]			
						-	\ep^{jlm}\big[\tfrac1m(c^\L_h)^{mk\aaa}+2(a^{(5)\L}_h)^{mkt\aaa}\big]	$	\\	
$	(k^\NR_{\si gpp})^{jklm}	$	&		$	-	\de^{jm}\de^{kl}\big[\tfrac2{m^2}(f^\L_h)^{t\aaa}\tfrac1m(m^{(5)\L}_{5h})^{tt\aaa}\big]			
						-	\de^{jm}\tfrac1m(m^{(5)\L}_h)^{kl\aaa}			
						+	\de^{kl}\big[\tfrac1{m^2}(d^\L_{\prt h})^{jm\aaa}+\tfrac1{2m}\ep^{jnr}(H^{(5)\L}_{\prt h})^{nrtm\aaa}\big]	$	\\	
			&	\hskip20pt	$	+	\de^{kl}\ep^{jmn}\big[\tfrac1{2m^2}(c^\L_h)^{nt\aaa}+\tfrac1m(a^{(5)\L}_h)^{ntt\aaa}\big]			
						+	\ep^{jmn}\tfrac1m(a^{(5)\L}_h)^{nkl\aaa}	$	\\	
			&	\hskip20pt	$	-	\half\Big[\Big(\de^{jk}\big[\tfrac1{m^2}(d^\L_{\prt h})^{lm\aaa}+\tfrac1{2m^2}\ep^{lnr}(g^\L_{\prt h})^{nrm\aaa}			
						+	\tfrac2m(b^{(5)\L}_{\prt h})^{tlm\aaa}+\tfrac1m(b^{(5)\L}_{\prt h})^{ltm\aaa}			
						+	\tfrac1{2m}\ep^{lnr}(H^{(5)\L}_{\prt h})^{nrtm\aaa}\big]	$	\\	
			&	\hskip40pt	$	+	\ep^{jkm}\big[\tfrac1{2m^2}(e^\L_h)^{l\aaa}+\tfrac1{2m^2}(c^\L_h)^{tl\aaa}			
						+	\tfrac1m(m^{(5)\L}_h)^{lt\aaa}+\tfrac1m(a^{(5)\L}_h)^{tlt\aaa}\big]			
						-	\ep^{jkn}\tfrac2m(H^{(5)\L}_{\prt h})^{tnlm\aaa}\Big)			
						+	(k\leftrightarrow l)\Big]	$	\\	[2pt]
\hline
\hline
\end{tabular}
\end{table*}

Finally,
the spin-dependent terms coupling with the gravitational acceleration $\vec g$
are found to be 
\bea
H_{\si g}&=&(k^{\NR}_{\si g})^{jk} \si^j g^k
+(k^{\NR}_{\si g p})^{jkl} \si^j p^k g^l 
\nn\\
&&
+(k^{\NR}_{\si g pp})^{jklm} \si^j p^k p^l g^m,
\label{eq:sig}
\eea
where we define 
\beq
(k^{\NR}_{\si g pp})^{jklm}=(k^{\NR}_{\si g pp})^{jlkm}.
\eeq
The three terms in Eq.\ \rf{eq:sig} are all position independent.
Each term contains a rotation-invariant component.
The first involves the dipole spin-gravity operator $\vec\si\cdot\vec g$,
governed by the trace $(k^{\NR}_{\si\ph p})_j{}^j$.
In the absence of backgrounds,
the possible appearance of this operator in the hamiltonian $H_0$ 
has been the subject of discussion in the literature
\cite{yo01},
but it is known to be absent 
when the physical Foldy-Wouthuysen representation is adopted
\cite{st05}.
Here,
despite the use of this representation,
the dipole spin-gravity operator nonetheless appears
because the general EFT provides additional contributions to $H$,
so its detection would represent a signal of new physics.
Note that the third term in $H_{\si g}$ incorporates 
a rotation-invariant component $\propto \vec p^{\,2} \vec\si\cdot\vec g$,
which corrects the dipole spin-gravity coupling at $O(p^2)$,
along with another rotation scalar 
$(\vec \si \cdot \vec p)(\vec g \cdot \vec p)$.
The second term also contains a rotation scalar
associated with the component $(k^{\NR}_{\si g p})^{jkl}$ 
proportional to $\ep^{jkl}$,
which acts to correct the size of the last term in $H_0$.
The first and third terms in $H_{\si g}$ 
are P and T odd,
while the second is P and T even. 

It is useful to collect 
all the rotation-invariant contributions $\ring H$ to $H$,
which gives 
\bea
\ring H &=& 
\frac{\vec{p}^{\,2}}{2m}-(m -(k^\NR_{\ph})) \vec{g}\cdot\vec{z}
+(k^{\NR}_{\si g})^\prime \vec \si \cdot \vec g
\nn\\
&&
+ (k^\NR_{gp})^\prime \vec p \cdot \vec g 
+ \big( \tfrac{3}{4m} +(k^{\NR}_{\si g p})^\prime \big) 
(\vec{\si}\times\vec{p})\cdot\vec{g}
\nn\\
&&
+\tfrac 12 (k^{\NR}_{\si\ph p})^\prime
(\vec \si \cdot \vec p \hskip 3pt \vec g \cdot \vec z
+\vec g \cdot \vec z \hskip 3pt \vec \si \cdot \vec p)
\nn\\
&&
-\big(\tfrac{3}{4m}-\tfrac 12 (k^\NR_{\ph pp})^\prime \big)
(\vec{p}^{\,2}\vec{g}\cdot\vec{z}
+\vec{g}\cdot\vec{z}~\vec{p}^{\,2})
\nn\\
&&
+(k^{\NR}_{\si g pp})^\prime \vec{p}^{\,2} \vec \si \cdot \vec g
+(k^{\NR}_{\si g pp})^{\prime\prime} 
(\vec \si \cdot \vec p) \hskip 3pt (\vec g \cdot \vec p) ,
\label{hrotinv}
\eea
where the correction terms are ordered by increasing powers of the 3-momentum.
The coefficients with primes denote 
suitably normalized irreducible representations 
of the rotation group obtained from the nonrelativistic coefficients 
in Eqs.\ \rf{eq:ph}--\rf{eq:sig}.
The expression \rf{hrotinv} for $\ring H$
is of interest for certain experimental applications,
in part because the rotation invariance ensures 
that all terms take the same form at leading order
when expressed either in the laboratory frame or the Sun-centered frame.
This implies,
for example,
no leading-order dependence on the local sidereal time or laboratory colatitude
in experimental signals for these terms. 
Note,
however,
that $\ring H$ can be modified by boosts,
including the boost associated with the revolution of the Earth about the Sun.
Note also that some of the rotation-invariant terms \rf{hrotinv} 
have been proposed in other contexts as phenomenological modifications 
to conventional fermion-gravity couplings
\cite{lo64,ni10}.
The present work reveals how these and other effects 
can arise in the EFT context.

The expressions
\rf{hzero}, \rf{eq:ph}, \rf{eq:siph}, \rf{eq:g}, and \rf{eq:sig}
are derived for the special gravitational potential 
$\ph = -\vec g \cdot \vec z$
associated with a uniform gravitational field $\vec g$.
The nonrelativistic coefficients appearing in these expressions
are therefore strictly defined only in this restricted scenario,
which makes challenging direct comparisons of their measurements 
with results from other types 
of laboratory experiments and astrophysical observations.
It is therefore crucial to report the results 
of any given data analysis also as measurements of coefficients 
in the linearized Lagrange density $\cL^\L_\ps$ 
in Table \ref{tab:linferm1}.

The explicit relationships between 
coefficients in the nonrelativistic hamiltonian $H$
and coefficients in the linearized Lagrange density $\cL^\L_\ps$ 
are provided in Table \ref{tab:Hamiltonian}.
The first column of the table lists the nonrelativistic coefficients
appearing in the hamiltonian \rf{hamiltonian},
and the second column provides their expressions
in terms of the linearized coefficients
appearing in Tables \ref{tab:linferm1} and Table \ref{tab:linferm2}.
These results reduce correctly to those of Ref.\ \cite{km13}
in the SME limit in Minkowski spacetime with appropriate metric signature.
Note that all types of nonrelativistic coefficients are generated
from the linearized EFT.
However,
to guarantee that all components of all nonrelativistic coefficients 
are nonzero requires extending our present treatment of the EFT 
to include derivative background couplings $Dk$.
For example,
the piece $\cL^{(5)\L}_{\ps \prt\prt h}$ in Table \ref{tab:linferm1} 
acquires distinct and complementary contributions from the EFT
arising either from the commutator $[D_\mu, D_\nu]$
or from the anticommutator $\{D_\mu, D_\nu\}$ of covariant derivatives.
The symmetries of the construction imply 
that the former occur for any background,
while that the latter are associated
only with nonzero derivative background couplings $DDk$.

\subsection{Flavor dependence}
\label{Species dependence}

Different experiments may use distinct particle species $w$,
and many individual experiments use more than one species.
It is therefore necessary to incorporate multiple fermion flavors 
in the analysis.
The Lagrange density for the EFT based on GR coupled to the SM
with all known flavors of fermions is presented in Ref.\ \cite{kl21}. 
The coefficients can depend on flavor,
which introduces further types of WEP violations
in addition to the anomalous spin-gravity effects discussed above. 
For the experiments studied in the present work,
it suffices to consider electrons, protons, neutrons, and muons,
which we denote by $w=e$, $p$, $n$, and $\mu$. 
For simplicity,
we disregard here possible flavor-mixing effects,
which require accompanying violations of the conservation
of electric charge, baryon number, or lepton number.
For instance, 
we exclude positron-proton-curvature couplings in the EFT.

The linearized Lagrange density $\cL^\L_\ps$ in Table \ref{tab:linferm1} 
and the nonrelativistic hamiltonian $H$ in Eq.\ \rf{hamiltonian}
can be used to describe 
the gravitational couplings of any given fermion $w$.
For practical applications
and to report experimental results,
the various coefficients can be labeled accordingly.
For example,
the nonrelativistic coefficients of interest are then denoted as
$(k^\NR_{\ph})_w$,
$(k^\NR_{\ph p})_w^j$,
$(k^\NR_{\ph pp})_w^{jk}$,
$(k^{\NR}_{\si\ph})_w^j$,
$(k^{\NR}_{\si\ph p})_w^{jk}$,
$(k^{\NR}_{\si\ph pp})_w^{jkl}$,
$(k^\NR_{g})_w^j$,
$(k^\NR_{gp})_w^{jk}$,
$(k^\NR_{gpp})_w^{jkl}$,
$(k^{\NR}_{\si g})_w^{jk}$,
$(k^{\NR}_{\si g p})_w^{jkl}$,
$(k^{\NR}_{\si g pp})_w^{jklm}$,
and the hamiltonian $H$ is written as $H_w$.
The WEP violations in the EFT are thus encoded in expressions 
as the $w$ dependence of background coefficients.

Many experiments involve atoms or ions,
which can be viewed as aggregates of fermions.
In what follows,
we treat these following standard techniques in the literature
\cite{kl99,kv18}.
First,
the Schmidt model
\cite{ts37,bw52}
is adopted as the basis for determining sensitivities of individual nuclei
to the various nonrelativistic coefficients for the nucleons $p$ and $n$.
The sensitivity of the full atom or ion 
to all coefficients with $w=e$, $p$, $n$
and for exotic atoms also $w=\mu$
can then be obtained using standard electron-shell methods 
and general symmetry properties of the system.

Some investigations are performed with antiparticles.
For nonrelativistic laboratory experiments,
this implies that the data analysis requires 
instead using the nonrelativistic antiparticle hamiltonian $H_\wb$ 
corresponding to the particle hamiltonian $H_w$.
In the EFT,
the particle and antiparticle for each species
are both encoded in a single quantum field.
As a result,
$H_w$ and $H_\wb$ are simultaneously generated
from the block diagonalization of the relativistic hamiltonian
associated with the 
linearized Lagrange density $\cL^\L_\ps$ in Table \ref{tab:linferm1}.
The nonrelativistic antiparticle hamiltonian $H_\wb$
is given by expressions of the same forms 
as Eqs.\ \rf{hzero}--\rf{eq:sig},
but the corresponding nonrelativistic coefficients
$(k^\NR_{\ph})_\wb$,
$(k^\NR_{\ph p})_\wb^j$, 
$(k^\NR_{\ph pp})_\wb^{jk}$,
$(k^\NR_{g})_\wb^j$,
$(k^\NR_{gp})_\wb^{jk}$,
$(k^\NR_{gpp})_\wb^{jkl}$,
$(k^{\NR}_{\si\ph})_\wb^j$,
$(k^{\NR}_{\si\ph p})_\wb^{jk}$,
$(k^{\NR}_{\si\ph pp})_\wb^{jkl}$,
$(k^{\NR}_{\si g})_\wb^{jk}$,
$(k^{\NR}_{\si g p})_\wb^{jkl}$,
$(k^{\NR}_{\si g pp})_\wb^{jklm}$
that appear in these equations 
involve different combinations of the linearized coefficients
than those given in Table \ref{tab:Hamiltonian}.

The explicit conversion between $H_w$ and $H_\wb$
can be implemented using the charge-conjugation operator C,
which interchanges particles and antiparticles.
Incorporating the opposite 4-momenta of particles and antiparticles,
this conversion can conveniently be described instead 
in terms of the CPT properties of the operators in $H_w$.
The results can then be interpreted to obtain the equivalent 
of Table \ref{tab:Hamiltonian} for antiparticles.
We find that the expressions for the antiparticle nonrelativistic coefficients
take the same form as those in Table \ref{tab:Hamiltonian}
up to sign changes in front of certain linearized coefficients.
For linearized coefficients that have either no subscript or a subscript $h$,
these sign changes occur for coefficients 
with an odd number of spacetime indices.
In contrast,
for linearized coefficients having a subscript $\prt h$, 
the sign changes appear for coefficients with an even number
of spacetime indices.
For example,
we find that the particle expression
\beq
(k^\NR_{\ph})_w = 2(m^{\prime\L})_w^{\aaa}-2(a^\L)_w^{t\aaa}
+2m(e^\L_h)_w^{t\aaa} + \ldots
\eeq
converts to
\beq
(k^\NR_{\ph})_\wb = 2(m^{\prime\L})_w^{\aaa}+2(a^\L)_w^{t\aaa}
-2m(e^\L_h)_w^{t\aaa} + \ldots
\eeq
for the corresponding antiparticle.
We emphasize that 
the particle and antiparticle nonrelativistic coefficients can differ
for each species,
but only one independent set of linearized coefficients exists per species
because terms in the linearized EFT 
simultaneously include both particles and antiparticles.

Comparative experiments on particles and antiparticles
are typically sensitive to differences
between nonrelativistic coefficients.
The sign changes in converting from particles to antiparticles
imply that taking the difference of nonrelativistic coefficients
either cancels or doubles the contributions from the linearized coefficients.
One example of relevance in what follows is 
the difference of nonrelativistic coefficients 
\bea
&&\hskip-18pt
\De(k^\NR_{\ph})_{\wb w} 
\equiv 
(k^\NR_{\ph})_\wb-(k^\NR_{\ph})_w
\nn\\
&&\hskip15pt
= 4(a^\L)_w^{t\aaa} -4m(e^\L_h)^{t\aaa}_w +4m^2(a^{(5)\L}_h)^{ttt\aaa}_w
\label{diffone}
\eea
that involves the spin-independent $\ph$-coupled pieces 
of the particle and antiparticle hamiltonians.
Another is the difference 
of nonrelativistic coefficients 
\bea
&&\hskip-18pt
\De (k^\NR_{\si \ph})^{j}_{\wb w}
\equiv
(k^\NR_{\si \ph})^{j}_{\wb}-(k^\NR_{\si \ph})^{j}_w
\nn\\
&&\hskip0pt
= 4(b^\L)^{j\aaa}_w -2m\ep^{jkl}(g^\L_h)^{klt\aaa}_w 
+4m^2(b^{(5)\L}_h)^{jtt\aaa}_w
\label{difftwo}
\eea
that involves the spin-dependent $\ph$-coupled pieces 
of the particle and antiparticle hamiltonians.

\section{Potential differences}
\label{Potential differences}

The unconventional contributions 
to the linearized Lagrange density $\cL^\L_\ps$ in Table \ref{tab:linferm1}
and to the nonrelativistic hamiltonian $H$ in Eq.\ \rf{hamiltonian}
produce physical effects on the behavior of particles
studied in laboratory experiments and astrophysical observations.
Among the effects are dependences on the magnitudes and directions
of the particle momentum and spin,
on the particle flavor,
and on the gravitational potential
and the magnitudes and directions of its derivatives.
These dependences can be used in experiments  
designed to disentangle and measure 
the various coefficients for different species.
In this section,
we focus on the position dependence arising from the gravitational potential
and deduce constraints 
on the linearized coefficients appearing in Table \ref{tab:linferm1}
by comparing published measurements obtained at different potentials.
The results selected for analysis here
are chosen from among the numerous existing ones 
\cite{tables} 
to yield sharp constraints.
The values adopted are taken from Refs.\ 
\cite{%
kl99,bh08,hnl03,sf15,jb10,fa14,cs17,sp12,br14a,br14b,fc04,cs19,%
pw06,hb17,ms11,fr17,tc89,hm07,bb14,al16,jp85,cl05,%
fgt17,mh11,cb95,fr12,kt13,ba06,ba07a,ba07b,ba08,ba21}
for electrons, protons, and neutrons 
and Refs.\
\cite{jb79,bkl00,vwh01,gwb08,mdcpt,nowt16,ahm17}
for muons. 

Many experiments have already been performed
to measure SME coefficients for Lorentz violation
under the assumption that spacetime is Minkowski
\cite{tables}.
However,
the locations of the laboratories performing these experiments
are typically at different elevations
and hence at different gravitational potentials $\ph$.
Since the linearization \rf{asy} of a breve coefficient $\breve{k}^{\cdots}$
contains $h_{\mu\nu}$,
which depends on $\ph$ via 
\beq
h_{00} \approx -2\ph,
\quad
h_{0j} \approx 0,
\quad
h_{jk} \approx -2\ph \de_{jk},
\label{happrox}
\eeq
it follows that experiments at distinct laboratories 
purportedly measuring a given coefficient 
$\ol{k}_{\rm expt}^{\cdots}$
may in fact be measuring quantities that differ slightly
due to the gravitational coupling,
\bea
\ol{k}_{\rm expt}^{\cdots} 
&=&
\ol{k}_\eff^{\cdots} + (k^\L)^{\cdots\mu\nu} h_{\mu\nu}
\nn\\
&\approx&
\ol{k}_\eff^{\cdots} - 2(k^\L)^{\cdots \aaa}\ph.
\label{expt}
\eea
Comparing experiments measuring coefficients $\ol{k}_{\rm expt}^{\cdots}$ 
at different elevations can therefore provide access 
to the combination $(k^\L)^{\cdots \aaa}$ of linearized coefficients.

We note in passing that the expression \rf{expt} depends
on the absolute value of the gravitational potential $\ph$.
However,
the comparison of coefficients $\ol{k}_{\rm expt}^{\cdots}$
at two different points $\vec x_1$ and $\vec x_2$
involves only the potential difference 
$\De\ph = \ph(\vec x_2) - \ph(\vec x_1)$,
and so the zero of the potential is irrelevant.
The dependence of observables on $\De \ph$ rather than $\ph$
is conventionally associated with gauge invariance,
but here it is an artifact of the linearization procedure 
and holds true despite the presence of gauge-violating terms 
in the Lagrange density in Table \ref{tab:linferm1}.

More generally,
the absolute value of $\ph$ can become an observable
in the presence of gauge-violating terms from beyond-Riemann gravity,
so sufficiently precise experiments could in principle measure it.
This would require a treatment including higher orders in $h_{\mu\nu}$,
and for some applications would also involve a reformulation of the procedure
to account for fluctuations around a cosmological spacetime 
rather than the approximately Minkowski spacetime considered here.
The measured coefficients $\ol{k}_{\rm expt}^{\cdots}$
would then have the schematic dependence
$\ol{k}_{\rm expt} \sim \ol{k}_\eff + k^\L h + k^{\rm Q} h h + \ldots$,
so comparing experimental results 
could permit measurements of the combinations $k^{\rm Q} \ph$,
ultimately leading to measurement of the absolute value of $\ph$
provided at least one coefficient $k^Q$ is nonzero. 
Developing this line of investigation is of definite interest
and would become vital in the event 
of a compelling nonzero experimental signal,
but it lies beyond our present scope.

\renewcommand\arraystretch{1.6}
\begin{table}
\caption{
\label{tab:elevations}
Laboratory elevations.} 
\setlength{\tabcolsep}{5pt}
\begin{tabular}{lcl}
\hline
\hline
Laboratory location & Elevation (m) & Ref. \\
\hline
Amherst, MA, USA	&	70	&	\cite{cb95}, \cite{sp12}	\\
Bad Homburg, Germany	&	165	&	\cite{fgt17}	\\
Berkeley, CA, USA	&	186	&	\cite{mh11}, \cite{al16}	\\
Berlin, Germany	&	30	&	\cite{hm07}	\\
Berlin, Germany	&	75	&	\cite{kt13}, \cite{fa14}, \cite{cs19}	\\
Boston, MA, USA	&	5	&	\cite{tc89}, \cite{fc04}	\\
Boulder, CO, USA	&	1637	&	\cite{jp85}, \cite{br14a}, \cite{br14b}	\\
Darmstadt, Germany	&	139	&	\cite{bb14}	\\
Geneva, Switzerland	&	442	&	\cite{cs17}, \cite{jb79}	\\
Heidelberg, Germany	&	309	&	\cite{cl05}	\\
Los Alamos, NM, USA	&	2226	&	\cite{vwh01}	\\
New York, NY, USA	&	24	&	\cite{gwb08}	\\
Paris, France	&	66	&	\cite{pw06}, \cite{hb17}	\\
Perth, Australia	&	14	&	\cite{hm07}	\\
Princeton, NJ, USA	&	37	&	\cite{jb10}, \cite{ms11}, \cite{fr17}	\\
Seattle, WA, USA	&	26	&	\cite{bh08}, \cite{fr12}	\\
Hsinchu, Taiwan	&	71	&	\cite{hnl03}	\\
\hline
\hline
\end{tabular}
\end{table}

\renewcommand\arraystretch{1.2}
\begin{table*}
\caption{
\label{tab:tildecoefficients}
Definitions for tilde combinations of linearized coefficients.} 
\setlength{\tabcolsep}{6pt}
\begin{tabular}{cl}
\hline
\hline
Coefficient & Combination \\
\hline
$	(\bt^\L)^{J\bbb}	$&$	(b^L)^{J\bbb} -\half\ep^{JKL}(H^L)^{KL\bbb}+m\left((d_h^L)^{JT\bbb}-\half\ep^{JKL}(g_h^L)^{KLT\bbb}\right)	$	\\	
$	(\bt^\L)^{T\bbb}	$&$	(b^L)^{T\bbb}-m(g_h^L)^{XYZ\bbb}	$	\\	
$	(\bt^\L)^{*J\bbb}	$&$	(b^L)^{J\bbb}+ \half \ep^{JKL}(H^L)^{KL\bbb}- m\left((d_h^L)^{JT\bbb} + \half \ep^{JKL}(g_h^L)^{KLT\bbb}\right)	$	\\	[4pt]
$	(\ct^\L)^{-\bbb}	$&$	m\left((c_h^L)^{XX\bbb}-(c_h^L)^{YY\bbb}\right)	$	\\	
$	(\ct^\L)^{Q\bbb}	$&$	m\left((c_h^L)^{XX\bbb}+(c_h^L)^{YY\bbb}-2(c_h^L)^{ZZ\bbb}\right)	$	\\	
$	(\ct^\L)^{J\bbb}	$&$	m \abs{\ep^{JKL}}(c_h^L)^{KL\bbb}	$	\\	
$	(\ct^\L)^{TJ\bbb}	$&$	m\left(c_h^L)^{TJ\bbb}+(c_h^L)^{JT\bbb}\right)	$	\\	
$	(\ct^\L)^{TT\bbb}	$&$	m (c_h^L)^{TT\bbb}	$	\\	[4pt]
$	(\dt^\L)^{\pm\bbb}	$&$	m\left((d_h^L)^{XX\bbb} \pm (d_h^L)^{YY\bbb}\right)	$	\\	
$	(\dt^\L)^{Q\bbb}	$&$	m\left((d_h^L)^{XX\bbb}+(d_h^L)^{YY\bbb}-2(d_h^L)^{ZZ\bbb}-(g_h^L)^{YZX\bbb}-(g_h^L)^{ZXY\bbb}+2(g_h^L)^{XYZ\bbb}\right)	$	\\	
$	(\dt^\L)^{XY\bbb}	$&$	m\left((d_h^L)^{XY\bbb}+(d_h^L)^{YX\bbb}-(g_h^L)^{ZXX\bbb}+(g_h^L)^{ZYY\bbb}\right)	$	\\	
$	(\dt^\L)^{Y\!Z\bbb}	$&$	m\left((d_h^L)^{YZ\bbb}+(d_h^L)^{ZY\bbb}-(g_h^L)^{XYY\bbb}+(g_h^L)^{XZZ\bbb}\right)	$	\\	
$	(\dt^\L)^{ZX\bbb}	$&$	m\left((d_h^L)^{ZX\bbb}+(d_h^L)^{XZ\bbb}-(g_h^L)^{YZZ\bbb}+(g_h^L)^{YXX\bbb}\right)	$	\\	
$	(\dt^\L)^{J\bbb}	$&$	m\left((d_h^L)^{TJ\bbb}+\half (d_h^L)^{JT\bbb}\right)+\frac{1}{4}\ep^{JKL}(H^L)^{KL\bbb}	$	\\	[4pt]
$	(\Ht^\L)^{XT\bbb}	$&$	(H^L)^{XT\bbb}-m\left((d_h^L)^{ZY\bbb}-(g_h^L)^{XTT\bbb}-(g_h^L)^{XYY\bbb}\right)	$	\\	
$	(\Ht^\L)^{YT\bbb}	$&$	(H^L)^{YT\bbb}-m\left((d_h^L)^{XZ\bbb}-(g_h^L)^{YTT\bbb}-(g_h^L)^{YZZ\bbb}\right)	$	\\	
$	(\Ht^\L)^{ZT\bbb}	$&$	(H^L)^{ZT\bbb}-m\left((d_h^L)^{YX\bbb}-(g_h^L)^{ZTT\bbb}-(g_h^L)^{ZXX\bbb}\right)	$	\\	[4pt]
$	(\gt^\L)^{T\bbb}	$&$	(b^L)^{T\bbb}+m\left((g_h^L)^{XYZ\bbb}-(g_h^L)^{YZX\bbb}-(g_h^L)^{ZXY\bbb}\right)	$	\\	
$	(\gt^\L)^{c\bbb}	$&$	m\left((g_h^L)^{XYZ\bbb}-(g_h^L)^{ZXY\bbb}\right)	$	\\	
$	(\gt^\L)^{Q\bbb}	$&$	m\left((g_h^L)^{XTX\bbb}+(g_h^L)^{YTY\bbb}-2(g_h^L)^{ZTZ\bbb}\right)	$	\\	
$	(\gt^\L)^{-\bbb}	$&$	m\left((g_h^L)^{XTX\bbb}-(g_h^L)^{YTY\bbb}\right)	$	\\	
$	(\gt^\L)^{TJ\bbb}	$&$	m\abs{\ep^{JKL}}(g_h^L)^{KTL\bbb}	$	\\	
$	(\gt^\L)^{JK\bbb}	$&$	m\left((g_h^L)^{JTT\bbb}+(g_h^L)^{JKK\bbb}\right),~ ({\rm no}\ K\ {\rm sum},~J \not = K)	$	\\	
$	(\gt^\L)^{DJ\bbb}	$&$	(b^L)^{J\bbb}+m\ep^{JKL}\left((g_h^L)^{KTL\bbb}+\half (g_h^L)^{KLT\bbb}\right)	$	\\	[2pt]
\hline
\end{tabular}
\end{table*}

For laboratory experiments on the Earth,
the assumption of a uniform gravitational field
implies that the comparison of coefficients
at two different elevations $z_1$ and $z_2$
involves the potential difference 
$\De\ph = \ph(z_2) - \ph(z_1) = g (z_2-z_1)$. 
Using the expression \rf{expt}
to extract constraints on $(k^\L)^{\cdots \aaa}$ 
from results obtained at a fixed latitude and longitude
then requires only knowledge of the relative elevations
of the experimental measurements.
However,
the measurements compared here are performed in laboratories
located at distinct points on the Earth's surface.
Extracting constraints therefore requires knowledge 
of the potential difference $\De\ph$ at different geographic locations,
which can be challenging to establish.
Indeed,
the accurate determination of the gravitational equipotentials 
at the Earth's surface is a famous and formidable problem in geodesy
\cite{ts14}.
Observations can be made from the ground or from satellites,
and relevant options for height measurements
include 
elevations taken relative to mean sea level
or vertical data based on a reference geoid.
Issues such as ocean topography and local density fluctuations
must also be incorporated for an exact treatment.
Here,
our goal is to obtain initial estimates of the sensitivities
to linearized coefficients 
that are implied by published experimental limits on Lorentz violation.
For this purpose,
it suffices to adopt the values of the laboratory elevations 
above mean sea level listed in Table \ref{tab:elevations},
from which $\De\ph$ and hence approximate constraints
on linearized coefficients can be deduced.
Future experimental analyses that incorporate detailed precision techniques
to determine relative elevations and hence $\De\ph$
can be expected to sharpen substantially 
the results reported in this work.

Published results from the various experiments considered here
are typically expressed in the Sun-centered frame
\cite{sunframe}
and reported using a standard set of tilde coefficients,
which are linear combinations of coefficients
naturally appearing in the nonrelativistic limit
and are defined in Minkowski spacetime with gravitational effects disregarded.
For minimal terms involving operators of mass dimension $d\leq 4$,
these standard tilde combinations are summarized 
in Table P48 of Ref.\ \cite{tables}.
Generalizations of some of these have been found that include
also coefficients controlling nonminimal operators with $d\geq 5$
\cite{gkv14,kv15,dk16,kv18,kl19}.
However,
in the present context with gravitational couplings,
the published results expressed in terms of standard tilde coefficients 
must be converted using Eq.\ \rf{expt}
into expressions involving the linearized coefficients 
in Table \ref{tab:linferm2} instead.
The relevant combinations of the latter
that appear in the analysis to follow
are defined in Table \ref{tab:tildecoefficients}.
Note that in this table $J$, $K$, $L$ range over the values $X$, $Y$, $Z$.
Each row of the table contains the generalized tilde coefficient
followed by its expression in terms of the linearized coefficients
appearing in Table \ref{tab:linferm2}.
The notation and definitions for the generalized tilde coefficients
parallel those for the standard tilde coefficients.
Differences include the addition of the indices $\bbb$
representing the sum over $TT$, $XX$, $YY$, and $ZZ$
that emerges from the expansion \rf{expt},
and sign changes arising from index positions
and the convention for the metric signature.

\renewcommand\arraystretch{1.2}
\begin{table*}
\caption{
\label{tab:constraints}
Constraints on tilde combinations of linearized coefficients
for electrons, protons, and neutrons.}
\setlength{\tabcolsep}{4.5pt}
\begin{tabular}{cllllll}
\hline
\hline
Coefficient & Electron &Ref. & Proton & Ref. & Neutron &Ref. \\
\hline
$	|(\bt^\L)^{X\bbb}|	$	&	$	<	3	\times	10^{	-15	}	$	GeV	&	\cite{hnl03},\cite{bh08}	&	$	<	8	\times	10^{	-16	}	$	GeV	&	\cite{jb10},\cite{kt13},\cite{sf15}*	&	$	<	6	\times	10^{	-19	}	$	GeV	&	\cite{jb10},\cite{fa14}	\\	
$	|(\bt^\L)^{Y\bbb}|	$	&	$	<	3	\times	10^{	-15	}	$	GeV	&	\cite{hnl03},\cite{bh08}	&	$	<	8	\times	10^{	-16	}	$	GeV	&	\cite{jb10},\cite{kt13},\cite{sf15}*	&	$	<	6	\times	10^{	-19	}	$	GeV	&	\cite{jb10},\cite{fa14}	\\	
$	|(\bt^\L)^{Z\bbb}|	$	&	$	<	7	\times	10^{	-14	}	$	GeV	&	\cite{hnl03},\cite{bh08}	&	$	<	2	\times	10^{	-11	}	$	GeV	&	\cite{sp12},\cite{cs17}	&	$	<	5	\times	10^{	-5	}	$	GeV	&	\cite{tetal14}	\\	
$	|(\bt^\L)^{T\bbb}|	$	&	$	<	6	\times	10^{	-2	}	$	GeV	&	\cite{bh08},\cite{br14a}*,\cite{br14b}*	&				\dash						&		&	$	<	6	\times	10^{	5	}	$	GeV	&	\cite{fc04},\cite{br14b}*	\\	
$	|(\bt^\L)^{*X\bbb}|	$	&				\dash						&		&				\dash						&		&				\dash						&		\\	
$	|(\bt^\L)^{*Y\bbb}|	$	&				\dash						&		&				\dash						&		&				\dash						&		\\	
$	|(\bt^\L)^{*Z\bbb}|	$	&				\dash						&		&				\dash						&		&				\dash						&		\\	[10pt]
$	|(\ct^\L)^{-\bbb}|	$	&	$	<	1	\times	10^{	-10	}	$	GeV	&	\cite{ba06}*,\cite{cs19}	&	$	<	4	\times	10^{	-9	}	$	GeV	&	\cite{pw06},\cite{ms11},\cite{fr17}*,\cite{hb17}*	&	$	<	1	\times	10^{	-13	}	$	GeV	&	\cite{kl99}*,\cite{tc89},\cite{ms11}	\\	
$	|(\ct^\L)^{Q\bbb}|	$	&	$	<	5	\times	10^{	-11	}	$	GeV	&	\cite{hm07},\cite{ba21}*	&	$	<	1	\times	10^{	3	}	$	GeV	&	\cite{pw06},\cite{bb14},\cite{hb17}*	&	$	<	1	\times	10^{	-5	}	$	GeV	&	\cite{ba08}*,\cite{al16}	\\	
$	|(\ct^\L)^{X\bbb}|	$	&	$	<	6	\times	10^{	-11	}	$	GeV	&	\cite{ba06}*,\cite{cs19}	&	$	<	3	\times	10^{	-9	}	$	GeV	&	\cite{pw06},\cite{ms11},\cite{fr17}*,\cite{hb17}*	&	$	<	3	\times	10^{	-13	}	$	GeV	&	\cite{kl99}*,\cite{jp85},\cite{ms11}	\\	
$	|(\ct^\L)^{Y\bbb}|	$	&	$	<	7	\times	10^{	-11	}	$	GeV	&	\cite{ba06}*,\cite{cs19}	&	$	<	9	\times	10^{	-10	}	$	GeV	&	\cite{pw06},\cite{ms11},\cite{fr17}*,\cite{hb17}*	&	$	<	3	\times	10^{	-13	}	$	GeV	&	\cite{kl99}*,\cite{jp85},\cite{ms11}	\\	
$	|(\ct^\L)^{Z\bbb}|	$	&	$	<	7	\times	10^{	-11	}	$	GeV	&	\cite{ba06}*,\cite{cs19}	&	$	<	2	\times	10^{	-9	}	$	GeV	&	\cite{pw06},\cite{ms11},\cite{fr17}*,\cite{hb17}*	&	$	<	1	\times	10^{	-13	}	$	GeV	&	\cite{kl99}*,\cite{tc89},\cite{ms11}	\\	
$	|(\ct^\L)^{TX\bbb}|	$	&	$	<	3	\times	10^{	-11	}	$	GeV	&	\cite{cs19},\cite{ba21}*	&	$	<	2	\times	10^{	5	}	$	GeV	&	\cite{cl05}*,\cite{pw06},\cite{hb17}*	&	$	<	1	\times	10^{	3	}	$	GeV	&	\cite{ba08}*,\cite{fgt17}*	\\	
$	|(\ct^\L)^{TY\bbb}|	$	&	$	<	1	\times	10^{	-11	}	$	GeV	&	\cite{cs19},\cite{ba21}*	&	$	<	2	\times	10^{	5	}	$	GeV	&	\cite{cl05}*,\cite{pw06},\cite{hb17}*	&	$	<	3	\times	10^{	3	}	$	GeV	&	\cite{ba08}*,\cite{fgt17}*	\\	
$	|(\ct^\L)^{TZ\bbb}|	$	&	$	<	3	\times	10^{	-11	}	$	GeV	&	\cite{ba06}*,\cite{cs19}	&	$	<	2	\times	10^{	5	}	$	GeV	&	\cite{cl05}*,\cite{pw06},\cite{hb17}*	&	$	<	3	\times	10^{	3	}	$	GeV	&	\cite{ba08}*,\cite{fgt17}*	\\	
$	|(\ct^\L)^{TT\bbb}|	$	&	$	<	1	\times	10^{	-10	}	$	GeV	&	\cite{ba07a}*,\cite{ba21}*	&	$	<	4	\times	10^{	8	}	$	GeV	&	\cite{pw06},\cite{mh11},\cite{hb17}*	&	$	<	7	\times	10^{	-3	}	$	GeV	&	\cite{cetal18}	\\	[10pt]
$	|(\dt^\L)^{+\bbb}|	$	&	$	<	6	\times	10^{	-10	}	$	GeV	&	\cite{ba07b}*,\cite{bh08}	&	$	<	2	\times	10^{	5	}	$	GeV	&	\cite{br14b}*,\cite{sf15}*	&				\dash						&		\\	
$	|(\dt^\L)^{-\bbb}|	$	&				\dash						&		&				\dash						&		&				\dash						&		\\	
$	|(\dt^\L)^{Q\bbb}|	$	&	$	<	7	\times	10^{	-10	}	$	GeV	&	\cite{ba07b}*,\cite{bh08}	&	$	<	6	\times	10^{	5	}	$	GeV	&	\cite{br14b}*,\cite{sf15}*	&				\dash						&		\\	
$	|(\dt^\L)^{XY\bbb}|	$	&	$	<	5	\times	10^{	-11	}	$	GeV	&	\cite{ba07b}*,\cite{bh08}	&				\dash						&		&				\dash						&		\\	
$	|(\dt^\L)^{Y\!Z\bbb}|	$	&				\dash						&		&				\dash						&		&				\dash						&		\\	
$	|(\dt^\L)^{ZX\bbb}|	$	&	$	<	5	\times	10^{	-10	}	$	GeV	&	\cite{ba07b}*,\cite{bh08}	&				\dash						&		&				\dash						&		\\	
$	|(\dt^\L)^{X\bbb}|	$	&	$	<	1	\times	10^{	-9	}	$	GeV	&	\cite{kl99}*,\cite{cb95},\cite{ba07b}*	&				\dash						&		&				\dash						&		\\	
$	|(\dt^\L)^{Y\bbb}|	$	&	$	<	1	\times	10^{	-10	}	$	GeV	&	\cite{kl99}*,\cite{cb95},\cite{ba07b}*	&				\dash						&		&				\dash						&		\\	
$	|(\dt^\L)^{Z\bbb}|	$	&				\dash						&		&	$	<	2	\times	10^{	-3	}	$	GeV	&	\cite{detal16}	&	$	<	4	\times	10^{	-2	}	$	GeV	&	\cite{tetal14}	\\	[10pt]
$	|(\Ht^\L)^{XT\bbb}|	$	&				\dash						&		&				\dash						&		&				\dash						&		\\	
$	|(\Ht^\L)^{YT\bbb}|	$	&				\dash						&		&				\dash						&		&				\dash						&		\\	
$	|(\Ht^\L)^{ZT\bbb}|	$	&				\dash						&		&				\dash						&		&				\dash						&		\\	[10pt]
$	|(\gt^\L)^{T\bbb}|	$	&				\dash						&		&				\dash						&		&				\dash						&		\\	
$	|(\gt^\L)^{c\bbb}|	$	&				\dash						&		&				\dash						&		&				\dash						&		\\	
$	|(\gt^\L)^{Q\bbb}|	$	&				\dash						&		&				\dash						&		&				\dash						&		\\	
$	|(\gt^\L)^{-\bbb}|	$	&				\dash						&		&				\dash						&		&				\dash						&		\\	
$	|(\gt^\L)^{TX\bbb}|	$	&				\dash						&		&				\dash						&		&				\dash						&		\\	
$	|(\gt^\L)^{TY\bbb}|	$	&				\dash						&		&				\dash						&		&				\dash						&		\\	
$	|(\gt^\L)^{TZ\bbb}|	$	&				\dash						&		&				\dash						&		&				\dash						&		\\	
$	|(\gt^\L)^{XY\bbb}|	$	&				\dash						&		&				\dash						&		&				\dash						&		\\	
$	|(\gt^\L)^{YX\bbb}|	$	&				\dash						&		&				\dash						&		&				\dash						&		\\	
$	|(\gt^\L)^{ZX\bbb}|	$	&				\dash						&		&				\dash						&		&				\dash						&		\\	
$	|(\gt^\L)^{XZ\bbb}|	$	&				\dash						&		&				\dash						&		&				\dash						&		\\	
$	|(\gt^\L)^{YZ\bbb}|	$	&				\dash						&		&				\dash						&		&				\dash						&		\\	
$	|(\gt^\L)^{ZY\bbb}|	$	&				\dash						&		&				\dash						&		&				\dash						&		\\	
$	|(\gt^\L)^{DX\bbb}|	$	&	$	<	2	\times	10^{	-8	}	$	GeV	&	\cite{kl99}*,\cite{cb95},\cite{fr12}*	&				\dash						&		&				\dash						&		\\	
$	|(\gt^\L)^{DY\bbb}|	$	&	$	<	2	\times	10^{	-8	}	$	GeV	&	\cite{kl99}*,\cite{cb95},\cite{fr12}*	&				\dash						&		&				\dash						&		\\	
$	|(\gt^\L)^{DZ\bbb}|	$	&				\dash						&		&	$	<	4	\times	10^{	-3	}	$	GeV	&	\cite{detal16}	&	$	<	2	\times	10^{	-2	}	$	GeV	&	\cite{tetal14}	\\	
\hline
\hline
\end{tabular}
\end{table*}

We remark in passing that the tilde coefficient $(\bt^\L)^{J\bbb}$ 
is proportional 
to the spin-dependent nonrelativistic coefficient $(k^\NR_{\si\ph})^{J}$,
\beq
(\bt^\L)^{J\bbb} = -\half (k^\NR_{\si\ph})^{J} .
\label{kbtilde}
\eeq
No other coefficient in Table \ref{tab:tildecoefficients}
enjoys a simple relationship like this.
The expression \rf{kbtilde} emerges as follows. 
In a uniform gravitational field,
restricting the nonrelativistic hamiltonian $H$ 
given by Eq.\ \rf{hamiltonian}
to terms without dependence on the 3-momentum $\vec p$
and without derivatives of the potential $\ph$
retains only the perturbative corrections
involving the product $(k^\NR_{\ph})\ph$ in $H_\ph$
and $(k^\NR_{\si\ph})^{J}\ph$ in $H_{\si\ph}$. 
The latter combination couples to the spin $\si^J$.
However,
in the nonrelativistic limit in Minkowski spacetime,
the coupling to $\si^J$ is governed 
by the standard tilde coefficient $\bt^{J}$.
In approximately flat spacetime,
this coefficient acquires a dependence on $\ph$ given by Eq.\ \rf{expt}.
Comparing this dependence to the product $(k^\NR_{\si\ph})^{J}\ph$ 
then reveals the relationship \rf{kbtilde}.
Note that a similar line of reasoning suggests that $k^\NR_{\ph}$ 
is related to a combination $(\at^\L)^{T\bbb}$ of coefficients,
which we can define as
\beq
(\at^\L)^{T\bbb} \equiv -\half k^\NR_{\ph} .
\label{katilde}
\eeq
The corresponding combination of coefficients 
does indeed appear in the nonrelativistic hamiltonian in Minkowski spacetime
\cite{kl99},
but it produces no measurable effects in that context
because it amounts to an unobservable redefinition of the zero of energy
or,
equivalently,
because it can be removed from the theory via field redefinitions
\cite{ck97}. 
The observability of $k^\NR_{\ph}$
is thus confirmed to be a consequence of the coupling
to the gravitational potential,
the presence of which restricts the applicability of field redefinitions
\cite{ak04}.

In addition to using published laboratory experiments
to deduce constraints from Eq.\ \rf{expt},
we can also consider astrophysical observations. 
These have been used by Altschul 
to deduce a variety of constraints in the absence of gravity 
\cite{ba06,ba07a,ba07b,ba08,ba21}.
To compare these with laboratory results via Eq.\ \rf{expt} 
requires knowledge of the difference $\De\ph$
between the gravitational potential on astrophysical scales
and the potential in the laboratory.
The astrophysical sources of interest here
include pulsars and supernova remnants within the Milky Way,
along with active galaxies, quasars, and blazars
within and outside the Virgo supercluster. 
These sources span a substantial range of distance scales,
so the relevant gravitational potentials are disparate.
Moreover,
some of the coefficient constraints are derived from multiple sources,
while some involve propagation across significant distances.
Establishing definitive values for the relevant gravitational potentials
is therefore challenging.
Here,
we note that contributions to the gravitational potential $\ph_{\star}$
on these astrophysical scales typically are of order 
$\ph_{\star}\simeq -5\times 10^{-6}$,
substantially exceeding the contributions
$\ph_{\oplus}$ from the Earth
and $\ph_{\odot}$ from the Sun
at the laboratory location,
$\ph_{\oplus}\simeq -7\times 10 ^{-10}$ 
and $\ph_{\odot}\simeq -1\times 10^{-8}$.
To place conservative bounds on coefficients
via comparisons using Eq.\ \rf{expt},
we can therefore adopt the value $\De \ph\simeq 1\times 10^{-8}$.
This corresponds to assuming cancellation 
of the contributions $\ph_{\star}$
at the astrophysical source and at the laboratory.
The cancellation is unlikely to be exact in reality,
so a detailed investigation of the potential difference 
between any given astrophysical source and an Earth-based laboratory 
could well lead to improvements of one or two orders of magnitude
on the conservative constraints derived here.

With the above framework in place,
using Eq.\ \rf{expt} to perform comparisons
among the various laboratory and astrophysical results
yields bounds on many of the tilde coefficients 
defined in Table \ref{tab:tildecoefficients}.
Table \ref{tab:constraints} displays constraints 
on these combinations of linearized coefficients 
in the electron, proton, and neutron sectors.
The first column of the table lists the tilde coefficients.
The second column contains the constraints
deduced for the tilde coefficients in the electron sector,
and the third column lists the references
from which the constraints are deduced.
The fourth and fifth columns contain
analogous information for the proton sector,
while the last two columns concern the neutron sector.

In Table \ref{tab:constraints}, 
all constraints accompanied by two or more references
are obtained by comparison of two published limits as described above.
Where three or more references are cited,
a combination of experimental results and theoretical analysis 
has been used to establish the two published limits
adopted in deducing our constraints.
References in the table with an asterisk denote works 
containing results deduced on theoretical grounds,
as opposed to direct experimental measurements.
A few constraints listed in the table
are accompanied by a single experimental reference
\cite{tetal14,detal16,cetal18},
and these are derived using techniques described
in later sections of the present paper.
Note that many coefficients are unconstrained to date
by potential-difference comparisons.
Relevant results from a single elevation are available for many of them
\cite{tables},
but interpretation in the present context
must await second measurements at other laboratories. 

In addition to independent results at different elevations,
future prospects for improving the constraints in Table \ref{tab:constraints}
could include the use of a network of time-synchronized clocks
to provide simultaneous monitoring 
for the corresponding potential-dependent effects
\cite{gnome,qs20}.
For example,
the Global Network of Optical Magnetometers 
to search for Exotic physics (GNOME) 
is geographically spread and encompasses multiple elevations
\cite{gnome}.
Another option is to use space-based clocks,
which offer several advantages in searches for Lorentz violation
\cite{space,kt11}.
Comparisons of clocks on a space platform 
to ones on the surface of the Earth 
involve larger potential differences 
than attainable in ground-based laboratories
and can therefore be expected to yield substantially improved sensitivities
to the linearized coefficients.

\renewcommand\arraystretch{1.2}
\begin{table}
\caption{
\label{tab:muons}
Constraints on linearized coefficients for muons.}
\setlength{\tabcolsep}{6pt}
\begin{tabular}{cll}
\hline
\hline
Coefficient & Constraint &Ref. \\
\hline
$	|(b^\L)^{X\bbb}|	$	&	$	<	2	\times	10^{	-10	}	$	GeV	&	\cite{vwh01},\cite{gwb08}	\\	
$	|(b^\L)^{Y\bbb}|	$	&	$	<	2	\times	10^{	-10	}	$	GeV	&	\cite{vwh01},\cite{gwb08}	\\	
$	|(b^\L)^{Z\bbb}|	$	&	$	<	6	\times	10^{	-9	}	$	GeV	&	\cite{jb79},\cite{bkl00}*,\cite{mdcpt},\cite{gwb08}	\\	[10pt]
$	|(c^\L)^{TT\bbb}|	$	&	$	<	9	\times	10^{	5	}	$	GeV	&	\cite{jb79},\cite{nowt16}*,\cite{ahm17}*	\\	
\hline
\hline
\end{tabular}
\end{table}

The muon sector offers another interesting source of
constraints on beyond-Riemann physics.
An analysis along the lines performed above 
for electrons, protons, and neutrons 
can be performed to obtain constraints on linearized coefficients for muons.
Table \ref{tab:muons} displays the results.
Each row of this table provides the relevant linearized coefficient
and the constraint obtained, 
followed by the references used in deducing it.
The table has comparatively few entries,
reflecting in part the paucity of measurements at different elevations.
The experiments cited in the table
involve both boosted and nonrelativistic muons,
so a nonrelativistic treatment in terms of the tilde coefficients
is impractical.
Instead,
constraints can be deduced on individual linearized coefficients,
as displayed in the table.
With the successful operation of the Fermilab $g-2$ experiment
\cite{abi21},
future improvements on these results can be envisaged.

The techniques adopted here to obtain constraints
on linearized coefficients for electrons, protons, neutrons, and muons
could in principle be extended to other species.
In many cases,
sufficient data are lacking to obtain results,
but substantial datasets are available for certain species
such as quarks and neutrinos
\cite{tables}.
However,
treating these species systematically 
requires consideration of flavor-changing effects 
and hence a extension of the theoretical framework presented here.
This line of investigation would be of definite interest
but lies beyond our present scope.

\section{Gravitational accelerations}
\label{Free-fall experiments}

The unconventional contributions 
to the linearized Lagrange density $\cL^\L_\ps$ in Table \ref{tab:linferm1}
can modify the acceleration experienced by a system 
in a uniform gravitational field.
Experiments comparing the gravitational accelerations of different systems
therefore offer the opportunity to measure the coefficients
appearing in the nonrelativistic hamiltonian \rf{hamiltonian}.

Consider first the comparatively simple modification 
of the gravitational acceleration
provided by the spin-dipole term with operator $\vec \si \cdot \vec g$
in the rotation-invariant hamiltonian \rf{hrotinv}.
This term is governed by the coefficient $(k^{\NR}_{\si g})^\prime_w$
and can be studied in experiments comparing
the spin-precession frequencies of different atomic species
\cite{ve92,yo96,ki17}
or via a spin-torsion pendulum
\cite{he08}. 
Constraints on 
$(k^{\NR}_{\si g})^\prime_w \equiv (k^{\NR}_{\si g})_{wj}{}^j/3$
for electrons, protons, and neutrons are tabulated in Ref.\ \cite{ki17} as
\beq
(k^{\NR}_{\si g})^\prime_e < 10,
\quad
(k^{\NR}_{\si g})^\prime_p < 2\times 10^5,
\quad
(k^{\NR}_{\si g})^\prime_n < 10^3.
\eeq
The implications of these constraints 
for the linearized coefficients appearing 
in the Lagrange density $\cL_\ps^\L$ given in Table \ref{tab:linferm1} 
can be seen from the correspondence provided in Table \ref{tab:Hamiltonian}.
Note that only the trace contributions from $(k^{\NR}_{\si g})^{jk}_w$
are relevant for $(k^{\NR}_{\si g})^\prime_w$,
so the terms proportional to $\ep^{jkl}$ in Table \ref{tab:Hamiltonian}
play no role.
Note also that the remaining linearized coefficients 
contained in $(k^{\NR}_{\si g})^\prime_w$
are otherwise unconstrained by the experiments considered in this work.

In the remainder of this section,
we consider comparisons of the free-fall properties of
Sr atoms
\cite{tetal14},
Rb atoms
\cite{detal16},
and antimatter
\cite{aegis,alphagrav,gbar,ces20}.
We generalize the techniques of Refs.\ \cite{kl99,kv18}
to analyze these types of experiments
and use existing results to extract constraints
on nonrelativistic coefficients.

Consider a generic atom of mass $m_\latom$
formed from $N_e$ electrons, $N_p$ protons, and $N_n$ neutrons.
The hamiltonian $H_\latom$ governing the gravitational acceleration 
of the atom
contains a conventional piece and a correction $\de H$ 
arising from the unconventional terms in Table \ref{tab:linferm1}
that can be expressed as a sum of the perturbations for each particle,
\beq
\de H= \sum_{N=1}^{N_e} \de H_{e,N}
+\sum_{N=1}^{N_p} \de H_{p,N}
+\sum_{N=1}^{N_n} \de H_{n,N}.
\eeq
In the nonrelativistic limit and a uniform gravitational field, 
each component hamiltonian $\de H_{w,N}$ involves 
the explicit forms \rf{eq:ph}-\rf{eq:sig} for the corresponding particle,
containing coefficients labeled with the appropriate flavor $w = e$, $p$, $n$.

In free fall,
the motion of each component particle $w$
can be separated into two parts,
the motion with the atom and the motion relative to the atom.
The positions $\vec z_w$ and the momenta $\vec p_w$ of the particles
can therefore be written as
\beq
\vec z_w = \vec z_w^\atom + \vec z_w^\rel,
\quad 
\vec p_w = \vec p_w^\atom + \vec p_w^\rel.
\eeq
In terms of the position $\vec z_\latom$ and momentum $\vec p_\latom$
of the atom,
\beq
\vec z_w^\atom = \vec z_\latom,
\quad
\frac {\vec p_w^\atom} {m_w} = \frac {\vec p_\latom} {m_\latom},
\eeq
where $m_w$ is the mass of particle $w$.
In the experiments considered here,
the motion of the atom can be taken along the laboratory $z$ axis,
so $\vec z_\latom = z_\latom \hat z$
and $\vec p_\latom = p_\latom \hat z$.
The size of the atom is much smaller than the distance traveled,
so $\vec z_w \approx \vec z_w^\atom = \vec z_\latom$.
Also,
the speed of the atom is of order $10^{-9}$,
so $\vec p_w^\atom$ is negligible and
$\vec p_w \approx \vec p_w^\rel$.
At leading order,
the hamiltonian $H_\latom$ therefore takes the form
\beq
H_\latom \approx \frac{p_\latom^{2}}{2m_\latom} + m_\latom g z_\latom
+\de H(\vec z_w^\atom, \vec p_w^\rel),
\eeq

To derive the effective gravitational acceleration of the atom,
we apply the Ehrenfest theorem on the atomic motion to obtain 
\bea
m_\latom \frac{d^2}{dt^2}\langle z_{\textrm{atom}}\rangle 
&=& \frac{d}{dt} \langle p_{\textrm{atom}} \rangle
= - i\langle [p_\latom, H_\latom] \rangle,
\nn\\
&=&
- m_\latom g - i\langle [p_\latom, \de H] \rangle,
\eea
where the expectation values are taken in the atomic state
and we use the identity $[\vec z_\latom, \vec p_w^\rel] \equiv 0$.
Since the component hamiltonians $H_g$ and $H_{\si g}$
in Eqs.\ \rf{eq:g} and \rf{eq:sig}
are independent of the position
and $[\vec p_\latom, \vec p_w^\rel] \equiv 0$,
the only corrections to the gravitational acceleration
arise from $H_\ph$ and $H_{\si\ph}$
in Eqs.\ \rf{eq:ph} and \rf{eq:siph}.
Moreover,
the parity symmetry of the relative motion
guarantees the vanishing of the expectation of odd powers of $\vec p_w^\rel$.
The operator correcting the free-fall gravitational acceleration of the atom
can therefore be taken as
\bea
- i[p_\latom, \de H] 
&=&
\sum_{w,N_w} \Big[
(k^\NR_{\ph})_w + (k^\NR_{\ph pp})^{jk}_w p^j_w p^k_w 
\nn\\
&&
+ (k^{\NR}_{\si\ph})_w^j \si^j_w
+(k^{\NR}_{\si\ph pp})_w^{jkl} \si^j_w p^k_w p^l_w \Big] g,
\qquad
\label{operator}
\eea
which sums over contributions from the $N_w$ particles of species $w$.
The first two terms on the right-hand side of this expression
are independent of spin,
so they can be neglected in experiments
comparing the gravitational acceleration of an atom in different spin states. 
The last two terms are spin dependent
and hence can be neglected in experiments involving unpolarized atoms.
Note that in typical atoms
the expectation values of the momentum squared are of order
\cite{kl99}
$\langle\vec{p}^{\, 2}\rangle_e \simeq10^{-11}$ GeV$^2$
and $\langle\vec{p}^{\, 2}\rangle_p 
\approx \langle\vec{p}^{\, 2}\rangle_n
\simeq10^{-2}$ GeV$^2$,
so the contributions from electrons to the terms quadratic in momenta
can be neglected in what follows.

To determine the expectation value of the operator \rf{operator},
suppose the atom is in the state $\ket{\al,F,m_F}$,
where $F$ is the quantum number for the total angular momentum 
and $m_F$ is the azimuthal quantum number.
We can then decompose the right-hand side of the operator \rf{operator}
into combinations of irreducible tensor operators
and evaluate the expectation values using the Wigner-Eckart theorem
\cite{we27}.
For a rank-$r$ tensor operator $T^{(r)}_q$ with $q = -r, \ldots r$,
the expectation value can be written in the form 
\bea
&&\hskip-15pt
\langle \alpha, F, m_F | T^{(r)}_q | \alpha, F, m_F \rangle \nn\\
&&\hskip20pt
= \frac{\langle F, m_F | r, q, F, m_F \rangle}
{\langle F, F | r, q, F, F \rangle}
\langle \alpha, F, F | T^{(r)}_q | \alpha, F, F \rangle,
\quad
\label{cgcoefficients}
\eea
where $\langle F, m_F | r, q, F, m_F \rangle$ 
and $\langle F, F | r, q, F, F \rangle$
are Clebsch-Gordan coefficients.
It follows that
$\vev{T^{(r)}_q}$ vanishes for $q\neq0$ or $r>2F$.

Inspection of Eq.\ \rf{operator} reveals
that it contains spin-dependent tensor operators with rank $1\leq r \leq 3$.
The $q=0$ components of these operators are
\bea
T^{(1)}_0 &\supset& 
\si^z,\hskip5pt \si^z p^j p^j ,\hskip5pt \si^j p^j p^z,
\nn\\
T^{(2)}_0 &\supset& 
(\si^x p^y-\si^y p^x) p^z,
\nn\\
T^{(3)}_0 &\supset& 
\si^z p^x p^x+\si^z p^y p^y
\nn\\
&&
+2\si^x p^xp^z+2\si^y p^y p^z-2\si^z p^zp^z. 
\label{tensorop}
\eea
Except for the rank-three case,
these operators already appear in the Minkowski-spacetime treatment
of clock-comparison experiments
\cite{kl99}. 
In any given experiment involving a specific atom,
one or more of these operators may have vanishing expectation value.
Any nonzero expectation values can be expected to
produce modifications of the gravitational acceleration.

The above analysis is performed in the standard laboratory frame,
which is noninertial. 
As described in Sec.\ \ref{Theory},
our focus here is on unconventional effects 
that preserve translation invariance in the Sun-centered frame
\cite{sunframe},
which over a time scale large compared to experimental data acquisition
can be taken as an approximately inertial frame.
The nonrelativistic coefficients appearing in the operator \rf{operator}
are therefore constant in the Sun-centered frame,
and hence in the noninertial laboratory frame
they appear to vary with the local sidereal time $T_\oplus$
and the laboratory colatitude $\ch$.
The explicit form of the coefficient dependence on time 
can be obtained by performing the rotation \rf{eq:rotation}
from the laboratory frame to the Sun-centered frame.
The structure of the operators \rf{tensorop}
then reveals that the measured gravitational accelerations
in experiments with atoms can display oscillations with sidereal time
at harmonics up to third order in the sidereal frequency $\om_\oplus$.

\subsection{Sr atoms}

We consider first an experiment 
\cite{tetal14}
performed to compare the gravitational accelerations 
of two isotopes of strontium atoms having different spins,
the spin-9/2 fermion $^{87}$Sr
and the spin-zero boson $^{88}$Sr. 
The experiment measured the gravitational accelerations 
via the delocalization of atomic matter waves in a vertical optical lattice.
The laboratory is located at colatitude $\ch\simeq46.2^\circ$.

For present purposes,
the atoms can be modeled using standard techniques
\cite{kl99,kv18}.
The electrons in both $^{87}$Sr and $^{88}$Sr form a closed shell.
In the Schmidt model
\cite{ts37,bw52},
the spin $I = 9/2$ of the $^{87}$Sr nucleus
is associated with an unpaired valence neutron,
while all nucleons in the $^{88}$Sr nucleus are paired.
Any spin-dependent effects on the gravitational response 
of the two isotopes can therefore be assigned
to the spin $I$ of the $^{87}$Sr nucleus.

The total angular momentum $F$ of $^{87}$Sr is $F=I=9/2$,
so the atomic states of $^{87}$Sr
can be denoted as $\ket{\al,I=9/2, m_I}$,
where $\al$ represents the radial part of the wavefunction
and $m_I=-I, -I+1, \ldots, I$
is the spin projection along $\hat z$.
The orbital angular momentum $L$ of the $^{87}$Sr nucleus 
is found to be $L=4$
\cite{metal71},
so we can identify $I=L+1/2$.
The expectation values of the irreducible tensor operators \rf{tensorop} 
in the state $\ket{\al,I,I}$
can then be evaluated as 
\bea
&&\langle\si^z\rangle=1,
\quad
\langle\si^zp^jp^j\rangle=\langle \vec{p}^{\, 2} \rangle,
\quad
\langle\si^jp^jp^z\rangle=\frac{1}{2L+3}\langle \vec{p}^{\, 2}\rangle,\nn\\
&&\langle T^{(2)}_0\rangle=0,
\quad
\langle T^{(3)}_0\rangle=\frac{2L}{2L+3}\langle \vec{p}^{\, 2}\rangle.
\label{expectationvalue}
\eea
Note that the rank-two tensor operator 
provides no contribution to the gravitational acceleration.

Combining the results \rf{expectationvalue} 
with the Clebsch-Gordan coefficients \rf{cgcoefficients}
enables calculation of the expectation values
of the operator \rf{operator} correcting the gravitational acceleration.
Working in the laboratory frame,
we find that the spin-dependent correction 
${g_{\rm Sr,\si}}$
to the effective gravitational acceleration 
${g_{\rm Sr}}$
of an $^{87}$Sr atom polarized in the state $\ket{\al,I,m_I}$
is given by 
\begin{widetext}
\bea
\frac{g_{\rm Sr,\si}}{g}&=& 
-\frac{m_I}{m_{\rm Sr}}\Big\{ \tfrac29 (k^\NR_{\si\ph})_n^z
+\big[ \tfrac2{27}(k^\NR_{\si\ph pp})_n^{zjj}
+\tfrac1{99}(k^\NR_{\si\ph pp})_n^{jjz} 
\big]\langle\vec{p}^{\, 2}\rangle_n 
\nn\\
&&
\hskip40pt
+\frac{20m_I^2-293}{6930}\big[
(k^\NR_{\si\ph pp})_n^{zxx}
+(k^\NR_{\si\ph pp})_n^{zyy}
+2(k^\NR_{\si\ph pp})_n^{xxz}
+2(k^\NR_{\si\ph pp})_n^{yyz} 
-2(k^\NR_{\si\ph pp})_n^{zzz}
\big]\langle\vec{p}^{\, 2}\rangle_n\Big\},
\nn\\
\label{eq:87sr}
\eea
\end{widetext}
where $m_{\rm Sr}=80.9$ GeV is the mass of the $^{87}$Sr atom,
and repeated $j$ indices indicate summation 
over the spatial coordinates $j=x,y,z$ in the laboratory frame.
Note that the identity \rf{indexid}
is used in deriving this result.
Note also that the appearance of nonrelativistic coefficients
only in the neutron sector is a consequence of adopting the Schmidt model.
A more detailed nuclear model for $^{87}$Sr 
might reveal also dependence on nonrelativistic coefficients
in the proton sector,
but attempting this lies beyond our present scope.

The nonrelativistic coefficients appearing in Eq.\ \rf{eq:87sr}
are expressed in the laboratory frame
and therefore oscillate with the local sidereal time $T_\oplus$.
The explicit dependence on $T_\oplus$
can be displayed by transforming to the Sun-centered frame
using the rotation \rf{eq:rotation}.
Binning measurements of the effective gravitational acceleration 
$g_{\rm Sr}$
in sidereal time could therefore provide 
a signal of effects beyond Riemann geometry.
The oscillations can contain up to third harmonics
of the Earth's sidereal frequency $\om_\oplus$,
and each harmonic contains information
about different combinations of coefficients.
Here,
for purposes of comparison with the reported results
\cite{tetal14},
we treat the experimental data as averaged over sidereal time. 
A reanalysis of the experimental data incorporating time-stamp information
would yield additional information and be of definite interest.

The experimental analysis in Ref.\ \cite{tetal14}
reported the measurement of a parameter $k=(0.5\pm 1.1)\times 10^{-7}$,
defined via a phenomenological correction 
to the gravitational potential of the form
$\ph(z) = (1 + \be + k m_I) g z$,
where $\be$ is a species-dependent constant.
This expression contains only a term linear in $m_I$,
whereas the result \rf{eq:87sr} contains also a cubic term in $m_I$.
Since the experimental measurement used unpolarized $^{87}$Sr atoms,
the cubic term can be weighted equally over $m_I$
and replaced with its linear approximation.
Performing the match yields a bound 
on a combination of nonrelativistic coefficients in the neutron sector.
Given the comparatively small size of 
the expectation value $\langle\vec{p}^{\, 2}\rangle_n$,
it is convenient and standard practice
\cite{tables}
to separate the bound into two pieces,
one assuming only $(k^\NR_{\si\ph})_n^J$ is nonzero
and the other assuming only
$(k^\NR_{\si\ph pp})_n^{JKL}$ is nonzero.
In the canonical Sun-centered frame,
we thereby find the constraints
\beq
\Big|(k^\NR_{\si\ph})_n^Z\Big|<1\times10^{-4}\textrm{ GeV}
\label{srone}
\eeq
and
\beq
\Big|(k^\NR_{\si\ph pp})_n^{ZJJ}-0.4(k^\NR_{\si\ph pp})_n^{ZZZ}\Big|
<5\times10^{-2}\textrm{ GeV}^{-1}
\label{srtwo}
\eeq
at the 95\% confidence level.
Here,
repeated $J$ indices denote summation 
over the spatial coordinates $J=X,Y,Z$ in the Sun-centered frame.

Using the expressions in Table \ref{tab:Hamiltonian},
the above constraints on nonrelativistic coefficients
can be converted into bounds on the linearized coefficients
appearing in Table \ref{tab:linferm2}.
These in turn imply constraints on the terms
in the Lagrange density given in Table \ref{tab:linferm1}.
We can also express the results in terms of
the tilde coefficients defined in Table \ref{tab:tildecoefficients}.
This yields the constraints displayed in Table \ref{tab:constraints} 
associated with Ref.\ \cite{tetal14}.
The sensitivities achieved are seen to be complementary to those
derived in Sec.\ \ref{Potential differences} 
from comparisons of data at different potentials.

\subsection{Rb atoms}

Next,
we turn to an experiment 
\cite{detal16}
comparing the gravitational accelerations
of $^{87}$Rb atoms with different projections $m_F$
of the total angular momentum $F$.
The experiment used an atom interferometer oriented vertically
to compare the gravitational accelerations of the hyperfine states
$\ket{F=1, m_F=+1}$ and $\ket{F=1, m_F=-1}$.
The laboratory is 
at colatitude $\ch\simeq59.4^\circ$.

The $^{87}$Rb atom has a single valence electron in the $5^2S_{1/2}$ level,
so the total electronic angular momentum is $J=1/2$
with orbital angular momentum $L=0$,
so $J=L+1/2$.
The nucleus has spin $I=3/2$ with orbital angular momenta $L=1$
\cite{metal75},
so $I = L+1/2$.
In the Schmidt model,
the nuclear properties are assigned to a single valence proton.
This is expected to be a comparatively accurate description for $^{87}$Rb 
because the nucleus contains 50 neutrons,
which is a magic number.

Since the angular momenta for the electrons and nucleus
are good quantum numbers,
we can express the atomic state as the tensor product
of two parts,
one for the valence electron and one for the Schmidt proton 
\cite{kl99,kv18}:
\beq
\ket{\alpha, F, m_F} =
\langle F, m_F | J, m_J, I, m_I \rangle
\ket{\alpha\pr, J, m_J} \ket{\alpha^{\prime\prime}, I, m_I},
\label{eq:decompose}
\eeq
where $\langle F, m_F | J, m_J, I, m_I \rangle$ 
is a Clebsch-Gordan coefficient
and $\alpha$, $\alpha\pr$, $\alpha^{\prime\prime}$ 
denote the radial dependences.
Both the valence electron and the Schmidt proton 
have total angular momentum $L+1/2$, 
so evaluation of the expectation values 
of the irreducible tensor operators \rf{tensorop} 
in the component wavefunctions
again yields results of the form \rf{expectationvalue}.
We see that the rank-two tensor operators 
in the electron and proton sectors
have no effect on the gravitational acceleration
due to the vanishing \rf{expectationvalue} of their expectation values,
while the rank-three tensor operators
have $r>2F$ and so according to the Wigner-Eckart theorem
cannot contribute either. 

Collecting the results and working in the laboratory frame,
we obtain the spin-dependent correction $g_{\rm Rb, \si}$
to the effective gravitational acceleration $g_{\rm Rb}$
of a $^{87}$Rb atom in the state with azimuthal quantum number $m_F$,
\bea
\frac{g_{\rm Rb, \si}}{g}&=& 
-\frac{m_F}{m_{\rm Rb}}\Big\{
\tfrac56(k^\NR_{\si\ph})_p^z-\tfrac12(k^\NR_{\si\ph})_e^z
\nn\\
&&
\hskip15pt
+\big[\tfrac5{18}(k^\NR_{\si\ph pp})_p^{zjj}
+\tfrac1{12}(k^\NR_{\si\ph pp})_p^{jjz}\big]
\langle\vec{p}^{\, 2}\rangle_p\Big\},
\qquad
\label{eq:rb87}
\eea
where $m_{\rm Rb}=80.9$ GeV is the mass of the $^{87}$Rb atom.
Repeated $j$ indices denote summation 
over the spatial coordinates $j=x,y,z$ in the laboratory frame,
and the identity \rf{indexid} has again been used.

By virtue of the Earth's rotation,
the nonrelativistic coefficients in the result \rf{eq:rb87}
vary harmonically with the local sidereal time $T_\oplus$.
Conversion to the Sun-centered frame can be implemented 
using the rotation \rf{eq:rotation}.
Extracting the maximum information
about the nonrelativistic coefficients in the Sun-centered frame
therefore requires measuring 
both the time-independent gravitational acceleration
and its variations with the Earth's sidereal frequency $\om_\oplus$.
For present purposes,
we view the published result as averaged over sidereal time.
A search for sidereal dependence in the experimental data
would permit measurements of additional nonrelativisitc coefficients
and be well worthwhile. 

The analysis in Ref.\ \cite{detal16} yielded a measurement of
the E\"{o}tv\"{o}s ratio 
\cite{re22}
$\et = (0.2\pm 1.2)\times 10^{-7}$. 
Using the result \rf{eq:rb87},
we find
\bea
\et &\equiv&
2 \frac{g_{\rm Rb}(m_F = -1) - g_{\rm Rb}(m_F = +1)} 
{g_{\rm Rb}(m_F = -1) + g_{\rm Rb}(m_F = +1)}
\nn\\ 
&\approx& 
2\frac{g_{\rm Rb, \si}(m_F = -1)}{g}
\eea
at leading order in nonrelativistic coefficients.
Matching to the experimental result provides a constraint.
Following standard procedure
\cite{tables},
we express the constraint
first under the assumption that only  
the coefficients $(k^\NR_{\si\ph})_w^J$ are nonzero,
and then assuming only $(k^\NR_{\si\ph pp})_w^{JKL}$ are nonzero.
Evaluated in the Sun-centered frame,
this gives
\beq
\Big|(k^\NR_{\si\ph})_p^Z-0.6(k^\NR_{\si\ph})_e^Z\Big|
<2\times10^{-5}\textrm{ GeV}
\label{rbone}
\eeq
and 
\beq
\Big|(k^\NR_{\si\ph pp})_p^{ZJJ}
+0.3(k^\NR_{\si\ph pp})_p^{JJZ}\Big|
<7\times10^{-3} \textrm{ GeV}^{-1}
\label{rbtwo}
\eeq
at the 95\% confidence level.
Repeated $J$ indices denote summation over spatial indices $J=X,Y,Z$
in the Sun-centered frame.

Note that these results from $^{87}$Rb 
involve nonrelativistic coefficients in the electron and proton sectors,
whereas those from $^{87}$Sr discussed in the previous subsection
involve coefficients in the neutron sector.
The two experiments are thus complementary
in their coverage of the coefficient space.
Also,
in parallel with the treatment of results from $^{87}$Sr,
the above constraints can be converted 
into bounds on linearized coefficients using Table \ref{tab:Hamiltonian}
and thereby on the terms in the Lagrange density 
given by Tables \ref{tab:linferm1} and \ref{tab:linferm2}.
Constraints on the tilde coefficients
defined in Table \ref{tab:tildecoefficients} can also be obtained,
and these are assigned to the entries for Ref.\ \cite{detal16}
listed in Table \ref{tab:constraints}.
The prospects are excellent for future improved measurements
of these spin-gravity couplings
using recent developments in Rb interferometry
\cite{as20,zh20}.

\subsection{Antimatter}

Another interesting option is to compare the gravitational accelerations
of matter and antimatter. 
Several experimental collaborations are developing tests
to compare the free fall of hydrogen H and antihydrogen $\ol{\rm H}$
\cite{aegis,alphagrav,gbar,ces20}.
On the theory side,
the CPT transformation is formally defined in Minkowski spacetime
\cite{cpt}
but can be extended operationally to the gravitational context
\cite{ak04},
and possible manifestations of CPT violation
include different gravitational responses of matter and antimatter. 
The dominant spin-independent effects
on the gravitational couplings of H and $\ol{\rm H}$
have been determined 
for spontaneous violations of local Lorentz and diffeomorphism symmetries
\cite{kt11,kv15}.
In some scenarios,
the effects cancel for H but add for $\ol{\rm H}$,
leading to measurable and potentially striking differences
between the gravitational accelerations of H and $\ol{\rm H}$.
In this subsection,
we use the techniques developed in the present work 
to provide a treatment of explicit violations for H and $\ol{\rm H}$,
including spin-gravity couplings.

Consider first H.
Since the nucleus is a single proton,
no relative motion occurs and so 
$\langle\vec{p}^{\, 2}\rangle_p = 0$.
The operator \rf{operator} correcting the gravitational acceleration 
can therefore be restricted to $p^j$-independent terms,
and in the laboratory frame
the only relevant irreducible operators are the identity and $\si^z$. 
The ground state has 
$J=1/2$ and $L=0$ for the electron and $I=1/2$, $L=0$ for the proton.
Working in the Zeeman limit
where the total angular momentum $F$ is a good quantum number,
we denote the atomic state as $\ket{\al,F,m_F}$
with $F=0$ or $F=1$.
In this state,
the effective gravitational acceleration of H
in the laboratory frame is found to be
\beq
\frac{g_{\, \rm H}}{g}=
1 - \frac 1 {m_{\rm H}} \sum_{w=e,p}
\big((k^\NR_\ph)_w + m_F(k^\NR_{\si\ph})^z_w \big),
\label{hydrogeng}
\eeq
where $m_{\rm H}\simeq 0.939$ GeV is the mass of the H atom.
This expression contains both spin-independent
and spin-dependent terms.

A similar derivation holds for $\ol{\rm H}$.
The coefficients in the operator \rf{operator} 
must now be replaced with those appropriate for antiparticles,
as described in Sec.\ \ref{Species dependence}.
In particular,
the coefficients of interest become 
$(k^\NR_{\ph})_\wb$ and
$(k^{\NR}_{\si\ph})_\wb^j$,
where $\wb$ denotes the antiparticles $\ol e \equiv e^+$ and $\ol p$.
The calculation otherwise proceeds as before,
yielding the effective gravitational acceleration of $\ol{\rm H}$
in the laboratory frame as
\beq
\frac{g_{\, \ol{\rm H}}}{g}=
1 - \frac 1 {m_{{\rm H}}} \sum_{\ol w = \ol e, \ol p}
\big((k^\NR_\ph)_{\ol w} + m_F(k^\NR_{\si\ph})^z_{\ol w} \big),
\label{antihydrogeng}
\eeq
where the mass $m_{\ol{\rm H}}$ of the $\ol{\rm H}$ atom
is taken as $m_{\rm H}$ at leading order. 

To parametrize the difference
between the gravitational accelerations of H and $\ol{\rm H}$,
we adopt the E\"{o}tv\"{o}s ratio
\cite{re22}
defined as
\beq
\et
\equiv
2 \frac{g_{\ol{\rm H}} - g_{\rm H}} 
{g_{\ol{\rm H}} + g_{\rm H}} .
\eeq
Applying the results \rf{hydrogeng} and \rf{antihydrogeng},
we find
\bea
\et
&=& 
- \frac 1 {m_{\rm H}} \sum_{w=e,p}
\big( (k^\NR_\ph)_{\ol w} - (k^\NR_\ph)_w 
\nn\\ 
&&
\hskip 50pt
+ (m_{F,\ol{\rm H}}(k^\NR_{\si\ph})^z_{\ol w} 
- m_{F,{\rm H}}(k^\NR_{\si\ph})^z_w 
\big)
\qquad
\label{hhbareotvos}
\eea
in the laboratory frame.
We see that comparisons of the free fall of H and $\ol{\rm H}$
in different hyperfine states can produce different results
for the relative gravitational accelerations.
In principle,
measurements of distinct combinations of coefficients
could thereby be obtained.

If the H and $\ol{\rm H}$ atoms are unpolarized,
the E\"{o}tv\"{o}s ratio contains only spin-independent terms,
reducing to
\bea
\et &=&
- \frac 1 {m_{\rm H}} 
\big( 
\De(k^\NR_{\ph})_{\ol{e} e} + \De(k^\NR_{\ph})_{\ol{p} p} 
\big)
\nn\\
&=& 
-\frac 4 {m_{\rm H}} 
\big( 
(a^\L)_e^{T\bbb} -m(e^\L_h)^{T\bbb}_e +m^2(a^{(5)\L}_h)^{TTT\bbb}_e
\nn\\
&& 
\hskip15pt
+ (a^\L)_p^{T\bbb} -m(e^\L_h)^{T\bbb}_p +m^2(a^{(5)\L}_h)^{TTT\bbb}_p
\big).
\qquad
\eea
In this derivation,
the result \rf{diffone} has been used.
Also, the linearized coefficients appearing here 
are expressed directly in the Sun-centered frame,
as they are all invariant under the rotation \rf{eq:rotation}.

More generally,
if both the H and the $\ol{\rm H}$ atoms 
are in the same state $\ket{\al,F,m_F}$,
then spin-gravity couplings contribute to the E\"{o}tv\"{o}s ratio as well.
The rotation \rf{eq:rotation} to the Sun-centered frame
then generates dependence on the local sidereal time $T_\oplus$
and the colatitude $\ch$ of the laboratory.
We find
\bea
\et &=&
-\frac 1 {m_{\rm H}} \sum_{w=e,p}
\big( 
\De(k^\NR_{\ph})_{\wb w} 
+ m_F \De(k^\NR_{\si \ph})^{z}_{\wb w}
\big)
\nn\\
&=&
-\frac 1 {m_{\rm H}} 
\big[
\De(k^\NR_{\ph})_{\ol{e} e} + \De(k^\NR_{\ph})_{\ol{p} p} 
\nn\\
&&
+ m_F \big( \De(k^\NR_{\si \ph})^{Z}_{\ol{e}e}
+ \De(k^\NR_{\si \ph})^{Z}_{\ol{p}p}\big) 
\cos\ch
\nn\\
&&
+ m_F \big( \De(k^\NR_{\si \ph})^{X}_{\ol{e}e}
+ \De(k^\NR_{\si \ph})^{X}_{\ol{p}p}\big) 
\sin\ch\cos\om_\oplus T_\oplus
\nn\\
&&
+ m_F \big( \De(k^\NR_{\si \ph})^{X}_{\ol{e}e}
+ \De(k^\NR_{\si \ph})^{X}_{\ol{p}p}\big) 
\sin\ch\sin\om_\oplus T_\oplus
\big],
\qquad
\eea
which involves zeroth and first harmonics
in the Earth's sidereal frequency $\om_\oplus$.
Substitution of the results \rf{diffone} and \rf{difftwo}
provides an expression in terms of linearized coefficients
appearing in Table \ref{tab:linferm2},
which could be used to place constraints
on the terms in the Lagrange density given in Table \ref{tab:linferm1}.

In the future,
techniques for manipulating antihydrogen 
may be extended to heavier antiatoms.
Antideuterium,
which has an antideuteron nucleus,
is expected to be stable
and so could provide another option
for comparing the gravitational accelerations of matter and antimatter.
Since the nucleons in deuterium undergo relative motion,
contributions to the gravitational acceleration
can be expected from all the operators in Eq.\ \rf{operator}. 
Comparing the gravitational accelerations 
of deuterium and antideuterium 
would therefore provide unique sensitivities
to electron, proton, and neutron coefficients
controlling matter-gravity and antimatter-gravity couplings.

\section{Gravitational phase shifts}
\label{Interferometer experiments}

At the quantum level,
the propagation of a nonrelativistic particle 
in a uniform gravitational field
can be described by a Schr\"odinger equation 
containing a term for the gravitational potential energy.
As a result,
coherently split de Broglie waves propagating at different heights 
are predicted to acquire a relative quantum phase shift.
In the present context,
the unconventional contributions 
to the linearized Lagrange density $\cL^\L_\ps$ in Table \ref{tab:linferm1}
generate extra terms in the nonrelativistic hamiltonian \rf{hamiltonian},
and these imply that a neutron propagating in a gravitational potential
undergoes an additional phase shift.
In this section,
we use results from interferometric experiments
measuring the gravitationally induced phase shift for neutrons 
\cite{cow75,st80,we88,li97,zetal00,hetal14}
to derive some constraints on nonrelativistic coefficients
in the neutron sector.

The original experiment by Colella, Overhauser, and Werner (COW)
\cite{cow75}
used Bragg diffraction in silicon crystals
to measure the relative phases between two branches of a coherent neutron beam
traversing paths at different heights.
The experiment involved unpolarized neutrons,
so the spin-dependent operators appearing in
the components $H_{\si\ph}$ and $H_{\si g}$ 
of the nonrelativistic hamiltonian \rf{hamiltonian} 
produce no effects.
The neutron velocities in the experiment were nonrelativistic,
so contributions from momentum-dependent operators
are suppressed and can be neglected.
Also,
the momentum-independent operators in the component $H_g$ 
represent a position-independent pure potential
and so cannot be measured in the COW experiment.
The only relevant nonrelativistic coefficient in this case 
is therefore $(k^\NR_\ph)_n$.
Inspection of Eq.\ \rf{eq:ph} shows 
that it acts to rescale the conventional gravitational potential.

These considerations imply that
the effective gravitational acceleration $g_n$ of the neutron
in the COW experiment can be written as
\beq
\frac {g_n}{g} =
1-\frac{(k^\NR_\ph)_n}{m_n},
\eeq
where $m_n=0.940$ GeV is the neutron mass.
This expression is derived in the laboratory frame,
but it is a rotation scalar 
and so is valid also in the Sun-centered frame.
Note that no sidereal effects appear.
The original experiment
measured the gravitational acceleration to an accuracy of 10\%,
which implies the estimated constraint 
$(k^\NR_\ph)_n<1\times10^{-1}$ GeV.
However,
more recent versions of the experiment
have reached an accuracy of about 1\%  
\cite{zetal00},
corresponding to the constraint
\beq
(k^\NR_\ph)_n<1\times10^{-2}\textrm{ GeV}.
\label{cowresult}
\eeq
The first row of Table \ref{tab:Hamiltonian}
reveals the implications of this result
for linearized coefficients in the Lagrange density 
$\cL^\L_\ps$ given in Table \ref{tab:linferm1}.
Note that this set of linearized coefficients
are unobservable in nongravitational experiments
because they can be removed from the Lagrange density
via field redefinitions
\cite{ak04}. 

The above analysis applies to experiments with unpolarized neutrons.
An interferometric experiment 
applying magnetic fields to split a beam of neutrons 
into two beams having opposite polarizations and moving along different paths
has been performed with a neutron spin-echo reflectometer (OffSpec)
using the ISIS Neutron and Muon Source at the Rutherford Appleton Laboratory
\cite{hetal14}.
This setup is sensitive to spin-dependent gravitational couplings as well.
The beam neutrons are nonrelativistic
with comparatively small momenta,
so we can analyze the experiment 
using momentum-independent terms 
in the nonrelativistic hamiltonian \rf{hamiltonian}.
The components $H_g$ and $H_{\si g}$ are position independent
and hence for fixed initial polarization
cannot affect the measured experimental observables.
It follows that we can proceed using the $2\times2$ matrix operator 
\beq
g_{\textrm{spin}}=
g\Big(I-\frac{(k^\NR_\ph)_n}{m_n}I
-\frac{(k^\NR_{\ph\si})_n^j}{m_n} \si^j\Big)
\eeq
to describe the gravitational acceleration in spin space.

To gain insight,
consider first a scenario with the magnetic field
along a direction $\hat{z}^\prime$ in the standard laboratory frame
and the initial neutron polarization 
along an orthogonal direction $\hat{x}^\prime$.
The initial state can then be written as 
$\ket{+}_\xx= (\ket{+}_\zz +\ket{-}_\zz)/{\sqrt{2}}$.
After passing through the interferometer,
the neutron is in the final state 
$(e^{i\ph_{+}}\ket{+}_\zz+e^{i\ph_{-}}\ket{-}_\zz)/{\sqrt{2}}$,
where $\ph_{+}$ and $\ph_{-}$ are $2\times2$ matrices
governing the phase changes in the interferometer. 
These phase matrices can be obtained 
by replacing $g$ in the original calculation with $g_{\rm spin}$.
The experiment measured the final state in the $\pm\hat{x}^\prime$ direction.
The amplitude $A_+$ for finding this state in the $+\hat{x}^\prime$ direction
is
\beq
A_+ = \tfrac{1}{\sqrt{2}}\big(\bra{+}_\zz+\bra{-}_\zz\big)
\cdot 
\tfrac{1}{\sqrt{2}}\big(e^{i\ph_{+}}\ket{+}_\zz+e^{i\ph_{-}}\ket{-}_\zz\big).
\eeq
In the OffSpec analysis,
the corresponding probability $P_+$ was assumed to have the form
$P_+ = (1+\cos{\De \ph_{\textrm{eff}}})/2$.
Calculation shows the effective gravitational acceleration 
in this scenario is 
$g_{n, x^\prime} = g\big(1-((k^\NR_\ph)_n+(k^\NR_{\si\ph})^\xx_n)/m_n\big)$.
Generalizing the above derivation, 
we find that the effective gravitational acceleration
$g_{n,\hat s^j}$
for a neutron beam initially polarized along direction $\hat s^j$
is
\beq
g_{n,\hat s^j} = 
g\Big[1-\frac{(k^\NR_\ph)_n}{m_n}
-\frac{(k^\NR_{\si\ph})^j_n \hat{s}^j}{m_n}\Big]
\eeq
in the laboratory frame.

In the OffSpec experiment,
the maximum deviation of $g_{n,\hat s^j}$ from $g$
was found to be 2.5\%.
We can therefore place the constraint
\beq
\Big|(k^\NR_\ph)_n+(k^\NR_{\si\ph})^j_n \hat{s}^j\Big|
<2.5\times10^{-2}\textrm{ GeV}
\label{offspecresult}
\eeq
on nonrelativistic coefficients in the laboratory frame.
This result includes both spin-dependent and spin-independent effects.
The implications for the linearized coefficients
in the Lagrange density $\cL^\L_\ps$ given in Table \ref{tab:linferm1}
can be found using the relationships in Table \ref{tab:Hamiltonian}.
Note that the coefficient $(k^\NR_\ph)_n$
is a scalar under the rotation \rf{eq:rotation}
and so remains unchanged when transformed to the Sun-centered frame.
However,
$(k^\NR_{\si\ph})^j_n$ is found to contain oscillations 
in the local sidereal time $T_\oplus$ at the Earth's sidereal frequency.
In the Sun-centered frame,
where the coefficient $(k^\NR_{\si\ph})^J_n$ is constant,
the oscillations are instead attributed 
to the rotation of the initial polarization $\hat{s}^J$ with the Earth.

Future experiments with the neutron spin-echo spectrometer
have considerable potential for exploring
the variety of other unconventional contributions
to spin-dependent gravitational effects
described by the nonrelativistic hamiltonian \rf{hamiltonian}.
For example,
one option might be to use horizontally split beams
and compare phase changes for different initial spin orientations.
These changes are sensitive at leading order 
to the coefficients $(k^\NR_{\si g})^{jk}$ appearing in Eq.\ \rf{eq:sig}.

\section{Gravitational Bound States}
\label{Bound-state Experiments}

The nonrelativistic vertical motion 
of a neutron placed above a mirror 
in a uniform gravitational field
is governed by a one-dimensional Schr\"odinger equation
with an infinite potential well.
The bound states $\ps_\kk$ of the system are Airy functions,
and the lowest eigenenergies $E_\kk$ are of order $10^{-21}$ GeV
\cite{al12}.
The presence of the unconventional contributions 
to the linearized Lagrange density $\cL^\L_\ps$ in Table \ref{tab:linferm1}
shifts the energy levels and the transition frequencies of this system. 
In this section,
we consider experiments performed to measure 
the quantum properties of bouncing neutrons
\cite{netal02,cetal18}
and derive some constraints from existing experimental results
on nonrelativistic coefficients in the neutron sector.
Our analysis complements existing studies of Lorentz violation
in this system 
\cite{me18,iwa19,xs20}.

\subsection{Critical heights}

Each neutron eigenenergy $E_\kk$ can be associated 
with a critical height $z_\kk>0$ above the mirror,
\beq
E_\kk = m_n g z_\kk ,
\label{eq:zn}
\eeq
where $g$ is the effective gravitational acceleration of the neutron 
and the mirror is taken to be located at $z=0$.
The experimental values of the first two critical heights 
have been measured 
\cite{netal02}
as $z_1^{{\rm exp}}=12.2\pm1.9$ $\mu$m
and $z_2^{{\rm exp}}=21.6\pm2.3$ $\mu$m.
With conventional gravitational couplings,
the theoretical values for these critical heights are 
$z_1^{{\rm th}}=13.7$ $\mu$m
and $z_2^{{\rm th}}=24.0$ $\mu$m.
In this subsection,
we determine the corrections to these theoretical values
for the nonrelativistic hamiltonian \rf{hamiltonian}
and use the experimental measurements 
to constrain nonrelativistic coefficients.

Since the components $H_g$ and $H_{\si g}$ of the hamiltonian \rf{hamiltonian}
are independent of position,
they cannot affect the critical heights $z_\kk$.
Also,
the neutron momenta are small,
so momentum-dependent terms in the hamiltonian can be omitted.
The corrections to the critical heights
are therefore governed by the perturbation
\beq
\de H = 
(k^\NR_\ph)_n \vec{g}\cdot\vec{z}
+(k^{\NR}_{\si\ph})_n^j \si^j \vec{g}\cdot\vec{z}.
\label{eq:deltaH}
\eeq
The first term is spin independent,
while the second term depends on the neutron polarization. 
This perturbation affects $z_\kk$ through changes 
both to the eigenenergies $E_\kk$
and to the effective gravitational acceleration $g$ of the neutron.

For the spin-independent term in Eq.\ \rf{eq:deltaH},
we can use nondegenerate perturbation theory.
Including the corrections to both $E_\kk$ and $g$,
Eq.\ \rf{eq:zn} is modified into
\beq
\big( m_n g -(k^\NR_\ph)_n g \big) z_\kk^{\textrm{spin-indep}}
=E_\kk-(k^\NR_\ph)_n g \langle z \rangle,
\label{eq:spinindep}
\eeq
where 
$E_\kk$ is the unperturbed energy,
$g$ is the unperturbed gravitational acceleration,
and $\langle z \rangle\equiv\langle\ps_\kk|z|\ps_\kk\rangle=2E_\kk/(3m_n g)$.

The neutron spin introduces a degeneracy in the unperturbed energy levels,
which is split by the perturbation $\de H$.
Treating the spin-dependent term in Eq.\ \rf{eq:deltaH}
therefore requires degenerate perturbation theory.
Diagonalization of the degenerate perturbation can be performed directly
by writing
\beq
(k^{\NR}_{\si\ph})_n^j \si^j
=\sqrt{\big[(k^{\NR}_{\si\ph})_n^j\big]^2} \si_{\hat{k}},
\eeq
where 
$\big[(k^\NR_{\si\ph})_n^j\big]^2=
\sum_j (k^\NR_{\si\ph})_n^j(k^\NR_{\si\ph})_n^j$
and $\si_{\hat{k}}$ is the spin operator 
in the $(k^{\NR}_{\si\ph})_n^j$ direction.
This modifies Eq.\ \rf{eq:zn} to the form 
\beq
\Big( m_n g \mp \sqrt{\big[(k^{\NR}_{\si\ph})_n^j\big]^2} g \Big)
z_\kk^{\textrm{spin-dep}}
=E_\kk \mp \sqrt{\big[(k^{\NR}_{\si\ph})_n^j\big]^2} g \langle z \rangle,
\label{eq:spindep}
\eeq
where the upper and lower signs are for neutrons 
with spins aligned along and opposite to 
the direction $(k^{\NR}_{\si\ph})_n^j$,
respectively.

Combining the results \rf{eq:spinindep} and \rf{eq:spindep}
reveals that the modified critical heights are given by 
\beq
z_\kk\pr = z_\kk \Bigg(1+\frac{(k^\NR_\ph)_n}{3m_n}
\pm\frac{\sqrt{[(k^\NR_{\si\ph})_n^j]^2}}{3m_n}\Bigg).
\label{eq:znprime}
\eeq
This expression is derived in the laboratory frame,
but the form of the result is observer-rotation independent
and hence is also valid for coefficients 
$(k^\NR_\ph)_n$ and $(k^\NR_{\si\ph})_n^J$ 
in the Sun-centered frame.
Comparing with the experimental results
\cite{netal02}
and taking as usual only one coefficient nonzero at a time, 
we can deduce the constraints 
\bea
\big|(k^\NR_\ph)_n\big|&<&8.2\times10^{-1}\textrm{ GeV},
\nn\\
\sqrt{\big[(k^\NR_{\si\ph})_n^J\big]^2}&<&5.4\times10^{-1}\textrm{ GeV}
\label{eq:heightbounds}
\eea
at the 95\% confidence level.
The second of these results is obtained 
from the standard deviation of $z_\kk$.

The expression \rf{eq:znprime} for the modified critical heights 
is frame independent in form
and so at first glance might seem to contain no sidereal variations,
despite the dependence of the coefficients
$(k^\NR_{\si\ph})_n^j$ on $T_\oplus$ 
arising from the rotation \rf{eq:rotation} to the Sun-centered frame.
However,
the $\pm$ signs in Eq.\ \rf{eq:znprime}
refer to spins aligned along or against the direction
of $(k^\NR_{\si\ph})_n^j$,
which rotates at the Earth's sidereal frequency $\om_\oplus$. 
As a result,
if the experiment involves neutrons of definite polarization 
in the laboratory frame,
the polarization along $(k^\NR_{\si\ph})_n^J$ 
rotates in the Sun-centered frame.
The measured value of $z_\kk$ therefore can vary with sidereal time 
with the first harmonic of $\om_\oplus$.
An experimental search for this sidereal dependence 
would be of definite interest.

\subsection{Transition frequencies}

The transition frequencies between different energy levels $E_\kk$
have also been measured experimentally
via resonance with acoustic oscillations
\cite{cetal18}.
Denoting the transition frequency between $E_\kkk$ and $E_\kk$ 
by $\nu_{\kk\kkk}$,
the experiment obtained the results 
$\nu_{13}^{\rm exp}=464.8\pm1.3$ Hz
and $\nu_{14}^{\rm exp}=649.8\pm1.8$ Hz.
Under the assumption of conventional gravitational couplings,
the theoretical values for these frequencies are 
$\nu_{13}^{\rm th}=463.0$ Hz
and $\nu_{14}^{\rm th}=647.2$ Hz.
Next,
we find the corrections to these frequencies
arising from the nonrelativistic hamiltonian \rf{hamiltonian}
and use the experimental results to place bounds
on nonrelativistic coefficients for the neutron.  

\begin{figure}[htp]
\centering
\includegraphics[width=0.48\textwidth]{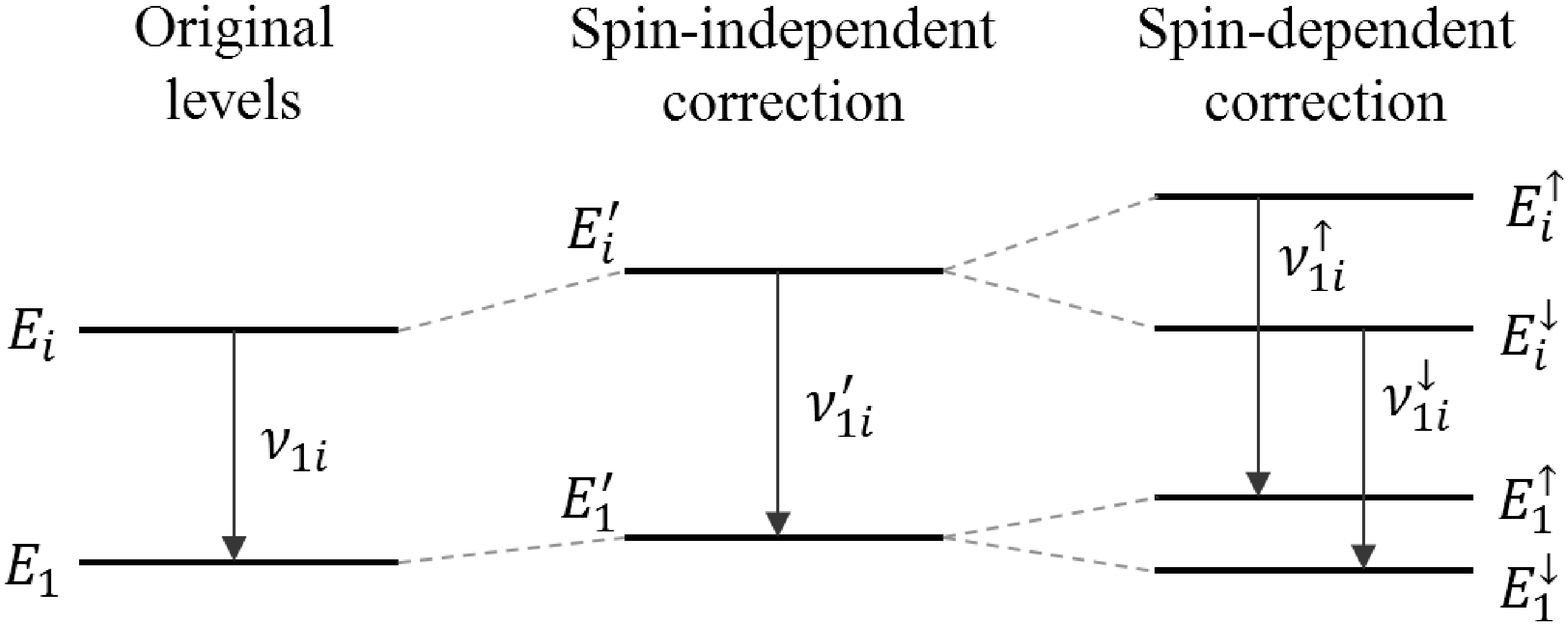}
\caption{Splitting of the neutron energy levels.}
\label{fig:splitting}
\end{figure}

The neutron momenta in the experiment are small,
so momentum-dependent terms in the hamiltonian \rf{hamiltonian}
can be neglected.
Moreover,
the term $(k^\NR_g)^jg^j$ in $H_g$ represents a constant potential
in this context
and hence leaves unaffected the energy differences.
The relevant terms in the perturbation hamiltonian are therefore
\beq
\de H = 
(k^\NR_\ph)_n \vec{g}\cdot\vec{z}
+(k^{\NR}_{\si\ph})_n^j \si^j \vec{g}\cdot\vec{z}
+(k^\NR_{\si g})^{jk}_n \si^j g^k.
\label{eq:RabiH}
\eeq
The first term is spin independent and shifts all energy levels,
while the others are spin dependent and split the energy levels.
The acoustic oscillations used in the experiment 
preserved the neutron spin,
so the experiment measured transitions between energy levels 
with same spin orientation,
as shown in Fig.\ \ref{fig:splitting}.

We use nondegenerate perturbation theory for spin-independent interactions
and degenerate perturbation theory for spin-dependent interactions.
After some calculation,
we find the energy shifts $\de E_\kk$ are given by 
\beq
\de E_\kk = -\tfrac23\frac{(k^\NR_\ph)_n}{m_n} E_\kk
\mp\sqrt{\Bigg(
\tfrac23\frac{(k^\NR_{\si\ph})_n^j}{m_n} E_\kk
+(k^\NR_{\si g})_n^{jz}g\Bigg)^2}
\label{eq:RabiEn}
\eeq
in the laboratory frame,
where the square inside the square root denotes summation over $j$,
$\sqrt{(k^j)^2}\equiv \big(\sum_j k^j k^j\big)^{1/2}$.
The upper and lower signs indicate neutrons with spins 
aligned along and opposite the direction 
$2(k^\NR_{\si\ph})_n^j E_\kk/(3m_n)+(k^\NR_{\si g})_n^{jz}g$,
respectively.
This direction typically differs for different energy levels 
because it depends on the unperturbed eigenenergies $E_\kk$.
As a result,
the spin-up state of the $\kk$-th energy level 
is oriented differently 
from the spin-up state of the $\kkk$-th energy level
when $\kk\neq\kkk$. 
The generic analysis of transitions between different energy levels
can therefore be involved.

For present purposes,
it suffices to adopt the standard practice
\cite{tables}
of taking only one of the coefficients 
$(k^\NR_{\si\ph})_n^j$ and $(k^\NR_{\si g})_n^{jk}$ 
to be nonzero at a time.
In this scenario,
the spins of either spin-up or spin-down states 
with different energy levels are aligned,
simplifying the discussion of transitions.
Also,
when only $(k^\NR_{\si g})_n^{jk}$ is nonzero,
different energy levels are split by the same amount.
This has no effect on the frequencies measured in the experiment,
so we can disregard $(k^\NR_{\si g})_n^{jk}$ in this context. 
Therefore,
assuming only one of $(k^\NR_{\si\ph})_n^j$ and $(k^\NR_{\si g})_n^{jk}$ 
is nonzero,
we find the energy differences in the laboratory frame 
are shifted according to
\beq
\de E_\kk-\de E_1=
- \tfrac23 \frac{(k^\NR_\ph)_n
\pm\sqrt{\big[(k^\NR_{\si\ph})_n^j\big]^2}}{m_n} (E_\kk-E_1).
\label{eq:RabidE}
\eeq

The form of the expression \rf{eq:RabidE}
is independent of rotations of the observer frame
and thus can be applied with the coefficients 
$(k^\NR_\ph)_n$ and $(k^\NR_{\si\ph})_n^J$ 
in the Sun-centered frame instead.
By comparing it to the experimental results
\cite{cetal18},
we deduce the constraints
\bea
\big|(k^\NR_\ph)_n\big|&<&1.3\times10^{-2}\textrm{ GeV},
\nn\\
\sqrt{\big[(k^\NR_{\si\ph})_n^J\big]^2}&<&7.8\times10^{-3}\textrm{ GeV}
\label{eq:qbounce}
\eea
at the 95\% confidence level.
The latter bound is derived
using the standard deviation of the transition frequencies.
Note that the constraints \rf{eq:qbounce}
are sharper than those in Eq.\ \rf{eq:heightbounds}
because transition frequencies can be measured 
more precisely than critical heights.
Using the appropriate rows in Table \ref{tab:Hamiltonian},
the above constraints can be converted
into conditions on linearized coefficients 
and hence on the terms in the Lagrange density 
given by Tables \ref{tab:linferm1} and \ref{tab:linferm2}.
We can also extract constraints on the tilde coefficients
introduced in Table \ref{tab:tildecoefficients}.
These are incorporated in Table \ref{tab:constraints}
as the entries associated with Ref.\ \cite{cetal18}.

In parallel with the result \rf{eq:znprime} for critical heights,
the expression \rf{eq:RabidE} for the transition frequencies
contains hidden dependence on the local sidereal time $T_\oplus$
emerging from the rotation \rf{eq:rotation} to the Sun-centered frame.
The $\pm$ signs represent spin projections along 
a direction determined by coefficients in the laboratory frame,
which rotates at the sidereal frequency
when expressed in the Sun-centered frame.
The measured values of the transition frequencies
can therefore fluctuate harmonically with $T_\oplus$
when polarized neutrons are used.
This signal would be worthwhile seeking in future experimental analyses.

\section{Summary}
\label{Discussion}

In this work,
we investigate observable effects arising in underlying theories
based on non-Riemann geometry or having a nongeometric basis,
and we constrain them by analyzing existing results
from laboratory experiments and astrophysical observations. 
The theoretical framework adopted for this purpose
is effective field theory based on GR coupled to the SM,
allowing for arbitrary backgrounds.
We focus on the LLI-EDV class of underlying theories,
which permit comparatively straightforward treatment 
of observable signals,
and consider primarily the effects of spin-gravity couplings
linearized around Minkowski spacetime.
Numerous first constraints are deduced on background coefficients
in these beyond-Riemann scenarios.

The methodology adopted for this work is described in Sec.\ \ref{Theory}.
The motivation and setup are presented 
for the class of underlying theories considered here,
with all fermion-gravity terms of mass dimension $d\leq 5$ 
in the linearized Lagrange density $\cL_\ps^\L$ 
displayed in Table \ref{tab:linferm1}.
The relationships between the linearized coefficients
appearing in this table and the underlying coefficients
in the full Lagrange density are listed in Table \ref{tab:linferm2}.
We use a generalized Foldy-Wouthuysen technique
to extract the corresponding nonrelativistic hamiltonian $H$,
with the explicit form for a uniform gravitational acceleration given in 
Eqs.\ \rf{hamiltonian}, \rf{hzero}, \rf{eq:ph}, 
\rf{eq:siph}, \rf{eq:g}, and \rf{eq:sig}.
The match between the nonrelativistic coefficients
appearing in $H$ and the linearized coefficients appearing in $\cL_\ps^\L$ 
is provided in Table \ref{tab:Hamiltonian}.
We also discuss the dependence of the coefficients
on particle and antiparticle flavor.

Using this methodology,
we explore the implications for the underlying theories
that arise from a variety of laboratory experiments 
and astrophysical observations.
We begin in Sec.\ \ref{Potential differences}
by considering constraints on linearized coefficients
that can be inferred from existing measurements
performed at different gravitational potentials. 
The generic dependence of a coefficient on the potential
is given in Eq.\ \rf{expt}.
Many of the experimental results in the literature 
turn out to be conveniently discussed 
in terms of a set of tilde coefficients,
defined in Table \ref{tab:tildecoefficients}.
The constraints obtained here
apply to the electron, proton, neutron, and muon sectors,
and they are summarized in 
Tables \ref{tab:constraints} and \ref{tab:muons}.

In Sec.\ \ref{Free-fall experiments},
we turn attention to experiments
comparing the gravitational accelerations of different atoms.
The modifications to the gravitational acceleration
relevant to these studies are given by the operator \rf{operator}.
Constraints from tests with $^{87}$Sr atoms of different spins
are derived and reported in Eqs.\ \rf{srone} and \rf{srtwo},
while those from tests with $^{87}$Rb atoms in different hyperfine states
are obtained in Eqs.\ \rf{rbone} and \rf{rbtwo}.
Future prospects are discussed for measurements 
of the gravitational acceleration of antimatter,
in particular for comparisons using H atoms and $\ol{\rm H}$ antiatoms.
Among the results is the derivation of the E\"{o}tv\"{o}s ratio
\rf{hhbareotvos}
describing the difference in free fall
between H and $\ol{\rm H}$ in various hyperfine states.

Studies of the quantum properties of nonrelativistic neutrons
also offer interesting sensitivity to fermion-gravity couplings.
In Sec.\ \ref{Interferometer experiments},
we examine interferometric experiments 
with split coherent neutron beams that traverse different paths
in a gravitational potential.
Constraints from the classic COW experiment with unpolarized neutrons
are derived in Eq.\ \rf{cowresult},
while ones from the spin-dependent OffSpec experiment  
are obtained in Eq.\ \rf{offspecresult}.
We also discuss measurements of the quantum bound states
of nonrelativistic neutrons above a neutron mirror.
Published results on the critical heights
for low-lying bound states lead to the constraints \rf{eq:heightbounds},
while measurements of transition frequencies
yield the bounds \rf{eq:qbounce}.
Where appropriate,
all our constraints on nonrelativistic coefficients
are translated into ones on tilde coefficients
and reported in Table \ref{tab:constraints}.

The methodology and results outlined in this work 
establish techniques for investigating gravitational effective field theories
arising from a class of underlying theories with beyond-Riemann structures.
The various calculations presented here illustrate the derivation 
of experimental and observational constraints for these theories.
The work establishes a path 
for further phenomenological and experimental studies
seeking unconventional signals 
in realistic gravitational effective field theories,
with considerable prospects for discovery.

\section*{Acknowledgments}

This work was supported in part
by the U.S.\ Department of Energy under grant {DE}-SC0010120
and by the Indiana University Center for Spacetime Symmetries.

\end{document}